\documentclass[10pt]{article}
\usepackage{amsmath,amssymb}
\usepackage{newtxtext,newtxmath,bm}
\usepackage{graphics,graphicx,epsfig}
\usepackage{natbib}
\usepackage[hmargin=1in,vmargin=1in]{geometry}
\numberwithin{equation}{section}

\begin{document}

%%%%%%%%%%%%%%%%%%%%%%%%%%%%%%%%%%%%%%%%%%%%%%%
%
\title{Continuum modeling of size-segregation and flow in dense, bidisperse granular media: Accounting for segregation driven by both pressure gradients and shear-strain-rate gradients}
\author{Harkirat Singh, Daren Liu, and David L. Henann\footnotemark[2]\\
School of Engineering, Brown University, Providence, RI 02912, USA}
\renewcommand*{\thefootnote}{\fnsymbol{footnote}}
\footnotetext[2]{Email address for correspondence: david\_henann@brown.edu}
\renewcommand*{\thefootnote}{\arabic{footnote}}
\date{}
\maketitle
\begin{abstract}
Dense mixtures of particles of varying size tend to segregate based on size during flow. Granular size-segregation plays an important role in many industrial and geophysical processes, but the development of coupled, continuum models capable of predicting the evolution of segregation dynamics and flow fields in dense granular media across different geometries has remained a longstanding challenge. One reason is because size-segregation stems from two driving forces: (1) pressure gradients and (2) shear-strain-rate gradients. Another reason is due to the challenge of integrating segregation models with rheological constitutive equations for dense granular flow. In this paper, we build upon our prior work, which combined a model for shear-strain-rate-gradient-driven segregation with a nonlocal continuum model for dense granular flow rheology, and append a model for pressure-gradient-driven segregation. We perform discrete element method (DEM) simulations of dense flow of bidisperse granular systems in two flow geometries, in which both segregation driving forces are present---namely, inclined plane flow and planar shear flow with gravity. Steady-state DEM data from inclined plane flow is used to determine the dimensionless material parameters in the pressure-gradient-driven segregation model for both spheres and disks. Then, the segregation model accounting for both driving forces is coupled with the nonlocal granular fluidity model---a nonlocal continuum model for dense granular flow---and predictions of the coupled, continuum model are tested against DEM simulation results across different cases of both inclined plane flow and planar shear flow with gravity, while varying parameters such as the size of the flow geometry, the driving conditions of flow, and the initial conditions. Overall, we find that it is crucial to account for both driving forces to capture segregation dynamics in dense, bidisperse granular media across both flow geometries with a single set of parameters.
\end{abstract}
%
%%%%%%%%%%%%%%%%%%%%%%%%%%%%%%%%%%%%%%%%%%%%%%%

\section{Introduction} \label{sec:intro}
When a granular medium is non-monodisperse, i.e., consisting of a mixture of particles of disparate size, grains tend to demix based on size during flow---a process referred to as size-segregation. Predicting the dynamics of size-segregation during flow is crucial in the design of various industrial processes where mixing is key and in understanding natural hazards, such as avalanches and landslides. The complex, spatially-inhomogeneous distributions in particle size that can arise during the segregation process \citep[e.g.,][]{savage88,gray1997,hill08,wiederseiner2011experimental,golick2009mixing,schlick2015,van2015underlying} pose a substantial challenge for modeling efforts, and significant effort over the last few decades has gone into developing continuum models for the evolution of particle-size distribution in a number of flow geometries (e.g., see the recent reviews of \citet{gray2018} and \citet{umbanhowar2019modeling}).

Here, we highlight two specific challenges for dense, granular media that motivate the present work. First, as per current understanding, there are two important driving forces for size-segregation in dense granular flows: (1) pressure gradients, typically arising due to gravity, and (2) shear-strain-rate gradients. In the presence of pressure gradients, the mechanisms of kinetic sieving and squeeze expulsion \citep{savage88,gray2018} result in a net flux of small particles to high-pressure regions and large particles to low-pressure regions, such as near a free surface. Pressure-gradient-driven segregation is widely recognized as a dominant driving force and is the focus of most size-segregation models in the literature \citep[e.g.,][]{gray05,gray06,gray11,marks2012,gajjar2014asymmetric,fan14,schlick2015,jones2018asymmetric,liu2019modeling,trewhela2021experimental,barker2021coupling,duan2021modelling}. Apart from pressure gradients, gradients in the shear-strain-rate can also drive size-segregation in dense flows, in which large particles segregate towards more rapidly shearing regions \citep{hill08,fan10,fan11a}, and comparatively fewer continuum modeling works have been dedicated to capturing this driving force \citep{fan11b,hill2014,liu_singh_arxiv}. While it is reasonable to expect segregation phenomenology in shallow free-surface flows to be predominantly driven by pressure gradients, the comparative contributions of the two driving forces is not always clear \citep{staron2015stress}, necessitating a model that accounts for both. 

The second challenge is coupling segregation modeling with rheology. The vast majority of the aforementioned segregation models do not involve rheological constitutive equations capable of predicting flow fields across different geometries. Instead, a certain flow field is assumed or measured from experiments or discrete element method (DEM) simulations and then prescribed as input to the segregation model. This is because modeling the rheological behavior of dense granular materials across flow geometries is a substantial challenge itself. The inertial, or $\mu(I)$, rheology \citep{midi2004,jop2005,dacruz2005,srivastava2021viscometric} is a common approach and uses dimensional arguments to relate the stress state to the state of strain rate at a point through a local constitutive equation. The $\mu(I)$ rheology works well in homogeneous shearing and certain other dense inertial flows, such as flow down an incline. However, the $\mu(I)$ rheology cannot capture a broad set of inhomogeneous flows spanning the quasi-static and dense inertial flow regimes, and a number of nonlocal continuum modeling approaches have been developed to capture the features of dense, inhomogeneous flows \citep{kamrin2019non}. 

Only a few works have sought to couple segregation modeling with rheological constitutive equations for a dense, bidisperse granular medium. For example, \citet{liu2019modeling} combined a rate-independent, Mohr-Coulomb-based, elasto-plasticity model with the pressure-gradient-driven segregation model of \citet{fan14} and applied the coupled model to dense, bidisperse flows in rotating drums and hoppers. More recently, \citet{barker2021coupling} combined a regularized version of the $\mu(I)$ rheology \citep{barker2017partial} with the pressure-gradient-driven segregation model of \citet{trewhela2021experimental} and applied the coupled model to dense, bidisperse flows down an inclined plane and in a rotating drum. Finally, in our recent work \citep{liu_singh_arxiv}, we generalized the nonlocal granular fluidity (NGF) model for dense granular flow \citep{kamrin2012,henann2013} from monodisperse to bidisperse granular systems and proposed a segregation model for the shear-strain-rate-gradient-driven flux. We showed that this coupled model could simultaneously capture segregation dynamics and flow fields in two different flow geometries in the absence of pressure gradients---vertical chute flow and annular shear flow. However, none of these models account for both pressure-gradient-driven and shear-strain-rate-gradient-driven segregation within a framework coupled to the rheological modeling of a dense granular medium. 

The purpose of this paper is to develop a model that integrates pressure-gradient-driven and shear-strain-rate-gradient-driven segregation, granular diffusion, and nonlocal rheological behavior spanning the quasi-static and dense inertial flow regimes into one framework capable of predicting segregation dynamics and flow fields. The starting point is our recent work \citep{liu_singh_arxiv}, which accounts for all but the pressure-gradient-driven flux. To this end, we supplement this model with a phenomenological constitutive equation for the pressure-gradient-driven flux, based on the works of \citet{gajjar2014asymmetric} and \citet{trewhela2021experimental}. To calibrate and test the coupled, continuum model, we perform DEM simulations using the open-source software LAMMPS \citep{lammps}. We return to the widely-studied, inclined plane flow geometry to calibrate the model for dense, bidisperse systems of both frictional spheres and disks, focusing on a fixed grain-size ratio. Then, we perform validation tests, in which we compare continuum model predictions of the steady-state flow fields and the transient evolution of the segregation dynamics against DEM simulation results in both inclined plane flow and an additional flow geometry not used in model calibration---planar shear flow with gravity. Both of these flow geometries involve both shear-strain-rate gradients and pressure gradients, so it is important to account for both driving forces of segregation to capture the segregation dynamics in both geometries with a single set of parameters.     

The remainder of this paper is organized as follows. First, we discuss the continuum framework for modeling coupled size-segregation and flow in Section~\ref{sec:conti_framework}. Specifically, we discuss the segregation model in Section~\ref{sec:seg_model} wherein we append the the constitutive equation for the pressure-gradient-driven flux \citep{gajjar2014asymmetric,trewhela2021experimental} to the model of \citet{liu_singh_arxiv}. To  determine the dimensionless material parameters that appear in the pressure-gradient-driven size-segregation model for both disks and spheres, we consider flows of bidisperse mixtures down an inclined plane in Section~\ref{sec:PGDS} and fit the model to steady-state DEM simulation results. Then, we perform validation tests of the continuum model against DEM simulation results for the transient evolution of the segregation dynamics, first for inclined plane flow in Section~\ref{sec:IPF_transient_comparisons} and second for an additional flow configuration not used in model calibration---planar flow with gravity---in Section~\ref{sec:PSwG_transient_comparisons} without parameter adjustment. In Section~\ref{sec:discussion}, we examine the relative contributions of the two driving forces of segregation in both flow geometries, and finally, we close with some concluding remarks in Section~\ref{sec:conclusion}.

\section{Continuum framework} \label{sec:conti_framework}
In this section, we discuss the continuum framework used to model dense, bidisperse granular mixtures. As in our prior work \citep{liu_singh_arxiv}, we utilize a mixture-theory-based approach \citep[e.g.,][]{gray2018,umbanhowar2019modeling}. While the continuum model developed in this work is a coupled model for size-segregation and flow, we primarily focus on the size-segregation aspect of the model in this section. The rheological constitutive equations that we utilize for bidisperse systems have been discussed in detail in our prior work \citep{liu_singh_arxiv} and for brevity, are only recapped in Appendix~\ref{app:flow_model}. Regarding notation, we use component notation, in which the components of vectors, $\boldsymbol{v}$, and tensors, $\boldsymbol{\sigma}$, relative to a set of Cartesian basis vectors $\{\boldsymbol{e}_i|i=1,2,3\}$ are denoted by $v_i$ and $\sigma_{ij}$, respectively. The Einstein summation convention is employed, and the Kronecker delta, $\delta_{ij}$, is utilized to denote the components of the identity tensor.

\subsection{Bidisperse systems}\label{subsec_mixt_seg}
We consider bidisperse granular systems made up of grains of two different sizes: large grains with average diameter $d^{\rm l}$ and small grains with average diameter $d^{\rm s}$. As in \citet{liu_singh_arxiv}, our focus is size-based segregation, and to eliminate density-based segregation, all grains are taken to be made of the same material with mass density $\rho_{\rm s}$. Both three-dimensional systems of bidisperse spheres and two-dimensional systems of bidisperse disks are considered in this work. The mass density $\rho_{\rm s}$ has units of mass per unit volume for spheres and mass per unit area for disks. For brevity, in what follows, we describe the continuum framework for three-dimensional systems of spheres, which may be straightforwardly adapted to two-dimensional systems of disks. In the mixture theory approach utilized here, several quantities are introduced for each species: the large grains and the small grains. Large-grain quantities are denoted with a superscript ${\rm l}$, and small-grain quantities with a superscript ${\rm s}$. The solid fractions, i.e., the volumes occupied by each species per unit total volume, are denoted as $\phi^{\rm l}$ and $\phi^{\rm s}$ for large and small grains, respectively, and the total solid fraction is $\phi = \phi^{\rm l} + \phi^{\rm s}$. The concentrations of each species are defined as $c^{\rm l} = \phi^{\rm l}/\phi$ and $c^{\rm s} = \phi^{\rm s}/\phi^{\rm s}$, so that $c^{\rm l} + c^{\rm s} =1$. The average grain size is defined as $\bar{d} = c^{\rm l} d^{\rm l} + c^{\rm s} d^{\rm s}$. 

\subsection{Kinematics of flow}\label{subsec_kine}
The velocity fields for each species are denoted as $v^{\rm l}_i$ and $v^{\rm s}_i$, and the mixture velocity field is defined as $v_i = c^{\rm l} v^{\rm l}_i + c^{\rm s} v^{\rm s}_i$. The strain-rate tensor for the mixture is defined as the symmetric part of the gradient of the mixture velocity: $D_{ij} = (1/2)(\partial v_i / \partial x_j + \partial v_j/ \partial x_i )$. In this work, we focus on dense flows, in which volume changes are small, and we make the common idealization that flow of the mixture proceeds at constant volume \citep[e.g.,][]{gray05,gray06,gray11,gajjar2014asymmetric,barker2021coupling,liu_singh_arxiv}. Under this idealization, the mixture velocity field is divergence-free, $\partial v_i/\partial x_i=0$; the strain-rate tensor is deviatoric, $D_{kk}=0$; and the total solid fraction $\phi$ remains constant. In this study, we use $\phi=0.6$ for spheres, and $\phi=0.8$ for disks. The equivalent shear strain rate is defined as $\dot{\gamma} = (2 D_{ij} D_{ij})^{1/2}$. 

\subsection{Balance of mass}\label{subsec_mass}
Using the mixture velocity field and the species-specific velocity fields, the relative volume fluxes for large and small grains are defined as $w^{\rm l}_i = c^{\rm l} (v^{\rm l}_i - v_i)$ and $w^{\rm s}_i = c^{\rm s} (v^{\rm s}_i - v_i)$, respectively, so that $w^{\rm l}_i + w^{\rm s}_i = 0_i$. Conservation of mass applied to the large grains requires that 
\begin{equation}\label{eq:cons_mass}
\frac{D c^{\rm l}}{D t} + \frac{\partial w^{\rm l}_i}{\partial x_i} = 0,
\end{equation}
where $D(\bullet)/Dt$ is the material time derivative. Similarly, conservation of mass applied to the small grains requires that $D c^{\rm s} /D t + \partial w^{\rm s}_i / \partial x_i = 0$, which, since $c^{\rm l}+c^{\rm s}=1$ and $w^{\rm l}_i + w^{\rm s}_i = 0_i$, is automatically satisfied when \eqref{eq:cons_mass} is satisfied. Therefore, in what follows, we utilize $c^{\rm l}$ as the field variable which describes the dynamics of size segregation. A constitutive equation is then required for the relative volume flux $w^{\rm l}_i$. In our previous work \citep{liu_singh_arxiv}, we proposed a constitutive equation for $w^{\rm l}_i$ that accounts for diffusion and shear-strain-rate-gradient-driven segregation. In Section~\ref{sec:seg_model}, we review this model and extend it to account for pressure-gradient-driven segregation. 

\subsection{Stress and the equations of motion}\label{subsec_eom}
The stress-related fields for the mixture are defined as follows: the symmetric Cauchy stress tensor $\sigma_{ij} = \sigma_{ji}$, the pressure $P =- (1/3)\sigma_{kk}$, the stress deviator $\sigma_{ij}' = \sigma_{ij} + P \delta_{ij}$, the equivalent shear stress $\tau = (\sigma_{ij}'\sigma_{ij}'/2)^{1/2}$, and the stress ratio $\mu=\tau/P$. The equations of motion are  
\begin{equation}\label{eq:EOM}
\phi\rho_{\rm s}\frac{D v_i}{Dt}= \frac{\partial \sigma_{ij}}{\partial x_j} + b_i,
\end{equation}
where $\phi$ is the constant total solid fraction, and $b_i$ is the non-inertial body force per unit volume (typically gravitational). Rheological constitutive equations are then required for the Cauchy stress $\sigma_{ij}$. In our previous work \citep{liu_singh_arxiv}, we generalized the nonlocal granular fluidity (NGF) model for dense granular flow from monodisperse to bidisperse systems. In this work, we continue to utilize the generalized NGF model to describe the rheological behavior of dense, bidisperse granular flows, which, for brevity, is summarized in Appendix~\ref{app:flow_model}.

\subsection{Segregation model}\label{sec:seg_model}
In this section, we discuss the constitutive equation for the relative volume flux $w^{\rm l}_i$. In dense granular flows, a bidisperse mixture tends to segregate due to both shear-strain-rate gradients and pressure gradients. Moreover, diffusion acts counter to segregation and tends to mix the two species. Accordingly, we take the relative volume flux to be comprised of three contributions as follows:
\begin{equation}\label{eq:flux_decomp}
w^{\rm l}_i = w^{\rm diff}_i + w^{\rm S}_i + w^{\rm P}_i,
\end{equation}
where $w^{\rm diff}_i$ is the diffusion flux, $w^{\rm S}_i$ is the shear-strain-rate-gradient-driven segregation flux, and $w^{\rm P}_i$ is the pressure-gradient-driven segregation flux. 

The first two contributions have been examined in detail in our prior work \citep{liu_singh_arxiv} and are briefly recapped here. First, we take the diffusion flux to be driven by concentration gradients as $w^{\rm diff}_i = -D \left(\partial c^{\rm l} / \partial x_i \right)$, where $D$ is the binary diffusion coefficient. In a dense, bidisperse granular mixture, the binary diffusion coefficient follows a well-established scaling \citep[e.g.,][]{trewhela2021experimental,barker2021coupling,duan2021modelling,liu_singh_arxiv} with the average grain size $\bar{d}$ and the shear strain rate $\dot{\gamma}$ as $D = C_{\rm diff} \bar{d}^2 \dot{\gamma}$, where $C_{\rm diff}$ is a dimensionless material parameter. Therefore, the diffusion flux is
\begin{equation}\label{eq:diff_eqn}
w^{\rm diff}_i = -C_{\rm diff} \bar{d}^2 \dot{\gamma} \frac{\partial c^l}{\partial x_i}.
\end{equation}
Second, in \citet{liu_singh_arxiv}, we isolated shear-strain-rate-gradient-driven segregation in two flow configurations with uniform pressure and proposed the following phenomenological constitutive equation for the flux $w^{\rm S}_i$:
\begin{equation}\label{eq:segS_eqn}
w^{\rm S}_i = C^{\rm S}_{\rm seg} \bar{d}^2 c^{\rm l}(1-c^{\rm l}) \frac{\partial \dot{\gamma}}{\partial x_i},
\end{equation}
where $C^{\rm S}_{\rm seg}$ is another dimensionless material parameter. In \eqref{eq:segS_eqn}, the pre-factor depends on the average grain size through $\bar{d}^2$ for dimensional consistency and involves a symmetric dependence on $c^{\rm l}$ through $c^{\rm l}(1-c^{\rm l})$, which ensures that segregation ceases when the bidisperse mixture becomes either all large ($c^{\rm l}=1$) or all small ($c^{\rm l}=0$) grains. While a symmetric dependence of the pre-factor on $c^{\rm l}$ was sufficient to capture the DEM data of \citet{liu_singh_arxiv} in the absence of pressure gradients, it contrasts with the asymmetric dependence that has been invoked to model pressure-gradient-driven size-segregation in the literature \citep[e.g.,][]{gajjar2014asymmetric,tunuguntla2017,jones2018asymmetric,trewhela2021experimental,barker2021coupling} and in the next paragraph of the present work. 

Next, we consider the segregation flux associated with pressure gradients, $w^{\rm P}_i$. We hypothesize that the segregation flux $w^{\rm P}_i$ is driven by gradients in the pressure $P$ and adopt the following phenomenological form for the constitutive equation for $w^{\rm P}_i$:
\begin{equation}\label{eq:segP_eqn}
w^{\rm P}_i = -C^{\rm P}_{\rm seg} \frac{\bar{d}^2 \dot{\gamma}}{P} c^{\rm l}(1-c^{\rm l}) (1- \alpha + \alpha c^{\rm l}) \frac{\partial P}{\partial x_i},
\end{equation}
where $C^{\rm P}_{\rm seg}$ and $\alpha$ are dimensionless material parameters. The dependence of the pre-factor on $\dot{\gamma}c^{\rm l}(1-c^ {\rm l})$ ensures that \eqref{eq:segP_eqn} satisfies several minimal requirements---namely, that the pressure-gradient-driven segregation flux is zero when there is no flow ($\dot\gamma=0$) or when the mixture becomes either all large ($c^{\rm l}=1$) or all small ($c^{\rm l}=0$) grains. Next, the dependence of the pre-factor on $\bar{d}^2/P$ ensures dimensional consistency. Finally, in contrast to \eqref{eq:segS_eqn}, the dependence of \eqref{eq:segP_eqn} on the factor $(1 - \alpha + \alpha c^{\rm l})$ introduces an asymmetric dependence on $c^{\rm l}$ into the flux constitutive equation. 

Several comments on the flux constitutive equation \eqref{eq:segP_eqn} are warranted:
\begin{enumerate}
\item The dependence of \eqref{eq:segP_eqn} on $c^{\rm l}$ through the asymmetric flux function $f(c^{\rm l}) = c^{\rm l}(1-c^{\rm l}) (1 - \alpha + \alpha c^{\rm l})$ follows directly from the work of \citet{gajjar2014asymmetric}, where $\alpha \in [0,1]$ is a parameter that controls the amount of asymmetry (denoted as $\gamma$ in \citet{gajjar2014asymmetric}). As discussed in \citet{gajjar2014asymmetric}, the flux function $f(c^{\rm l})$ has the following properties: (1) for $\alpha=0$, the symmetric flux function is recovered, (2) for $\alpha \in (0,1]$, the function's maximum is skewed from $c^{\rm l}=0.5$ towards $c^{\rm l}=1$, (3) for $\alpha \, \in \, (0, 0.5]$, it is convex, and (4) for $\alpha \, \in \, (0.5, 1]$, it is non-convex with a single inflection point. Previously, a value of $\alpha=0.46$ was obtained in \citet{gajjar2014asymmetric} by fitting to the experiments of \citet{wiederseiner2011experimental}, and a value of $\alpha=0.89$ was determined by fitting to experiments in \citet{van2015underlying}. In what follows, we will estimate $\alpha$ along with $C^{\rm P}_{\rm seg}$ for both simulated, frictional spheres and disks by fitting to DEM simulations.  
\item We note that the scaling of the pressure-gradient-driven segregation flux \eqref{eq:segP_eqn} with $\bar{d}^2\dot{\gamma}/P$ is quite similar to the scaling proposed in \citet{trewhela2021experimental} and utilized in \citet{barker2021coupling}. As pointed out in these works, combining a segregation flux that is inversely proportional to the pressure with a pressure-independent diffusion flux enables a model to capture the attenuation of segregation with increasing pressure, which was observed in the annular shear experiments of \citet{golick2009mixing} and the discrete simulations of \citet{fry2018effect,fry2019diffusion}. However, one difference is that \citet{trewhela2021experimental} introduces a second term in the denominator to prevent a singularity when the pressure is equal to zero, such as at a free surface. This additional term in the denominator depends on the product of the magnitude of the pressure gradient and the average grain size $\bar{d}$ and is multiplied by an additional (small) dimensionless constant. Here, we follow a similar approach but directly add a small constant to the pressure field when dealing with free-surface flows in Section~\ref{sec:IPF_transient_comparisons}.
\item Finally, it has been well established that the pressure-gradient-driven segregation flux should depend on the grain-size ratio $d^{\rm l}/d^{\rm s}$ \citep[e.g.,][]{schlick2015,tunuguntla2017,jones2018asymmetric,duan2021modelling,trewhela2021experimental,barker2021coupling}. However, the aims of this paper are to combine pressure-gradient-driven and shear-strain-rate-gradient-driven segregation fluxes within a coupled model for size-segregation and flow and then to use this model to examine the interplay between the mechanisms. Therefore, we a consider a single grain-size ratio of $d^{\rm l}/d^{\rm s}=1.5$ throughout, and considering the dependence of \eqref{eq:segP_eqn} on $d^{\rm l}/d^{\rm s}$ is beyond the scope of this paper. We expect that the substantial progress in the literature to characterize the role of the grain-size ratio \citep[e.g.,][]{trewhela2021experimental} may be incorporated into \eqref{eq:segP_eqn}.
\end{enumerate}

Combining \eqref{eq:diff_eqn}, \eqref{eq:segS_eqn}, and \eqref{eq:segP_eqn} with the balance of mass equation \eqref{eq:cons_mass}, we obtain the following differential relation governing the dynamics of $c^{\rm l}$:
\begin{equation}\label{eq:segP_model_seg}
\dfrac{D{c}^{\rm l}}{Dt} + \dfrac{\partial}{\partial x_i} \left( -C_{\rm diff} \bar{d}^2 \dot{\gamma} \frac{\partial c^l}{\partial x_i} + C^{\rm S}_{\rm seg} \bar{d}^2 c^{\rm l}(1-c^{\rm l}) \frac{\partial \dot{\gamma}}{\partial x_i} - C^{\rm P}_{\rm seg} \frac{\bar{d}^2 \dot{\gamma}}{P} c^{\rm l}(1-c^{\rm l}) (1- \alpha + \alpha c^{\rm l}) \frac{\partial P}{\partial x_i}  \right) = 0.
\end{equation}
The material parameters associated with segregation model are the set $\{C_{\rm diff}, C^{\rm S}_{\rm seg}, C^{\rm P}_{\rm seg}, \alpha\}$. As discussed in \citet{liu_singh_arxiv}, the parameter $C_{\rm diff}$ may be determined from measurements of the mean square displacement during homogeneous simple shearing, and the parameter $C^{\rm S}_{\rm seg}$ may be determined by examining size-segregation in flows, in which the pressure field is spatially uniform. Based on the DEM simulations of \citet{liu_singh_arxiv}, these parameters are $\{C_{\rm diff}=0.045, C^{\rm S}_{\rm seg}=0.08\}$ for frictional spheres and $\{C_{\rm diff}=0.20, C^{\rm S}_{\rm seg}=0.23\}$ for frictional disks. The remaining material parameters $\{C^{\rm P}_{\rm seg}, \alpha\}$, which are associated with the pressure-gradient-driven segregation flux, will be determined for frictional spheres and disks in the following section.    

\section{Pressure-gradient-driven segregation flux} \label{sec:PGDS}
In this section, we evaluate the constitutive equation for the pressure-gradient-driven segregation flux \eqref{eq:segP_eqn} and estimate the material parameters associated with this segregation flux $\{C^{\rm P}_{\rm seg}, \alpha\}$. To this end, we consider flows of both three-dimensional systems of dense, frictional spheres and two-dimensional systems of dense, frictional disks down an inclined plane, using discrete-element method (DEM) simulations. Details of the simulated granular systems for both spheres and disks are given in Appendix~\ref{app:DEM}. Among the grain interaction properties, the inter-particle friction coefficient $\mu_{\rm surf}$ is expected to play the most important role \citep{kamrin2014}. Determining the dependence of segregation phenomenology on $\mu_{\rm surf}$ is beyond the scope of this paper, and we maintain $\mu_{\rm surf}=0.4$ throughout for both spheres and disks. We note that the inclined plane flow configuration has been widely utilized in the literature \citep[e.g.,][]{savage88,gray05,gray06,wiederseiner2011experimental,marks2012,gajjar2014asymmetric,staron2015stress,barker2021coupling} to study size-segregation in dense granular materials. 

\begin{figure}[!t]
\begin{center}
\includegraphics{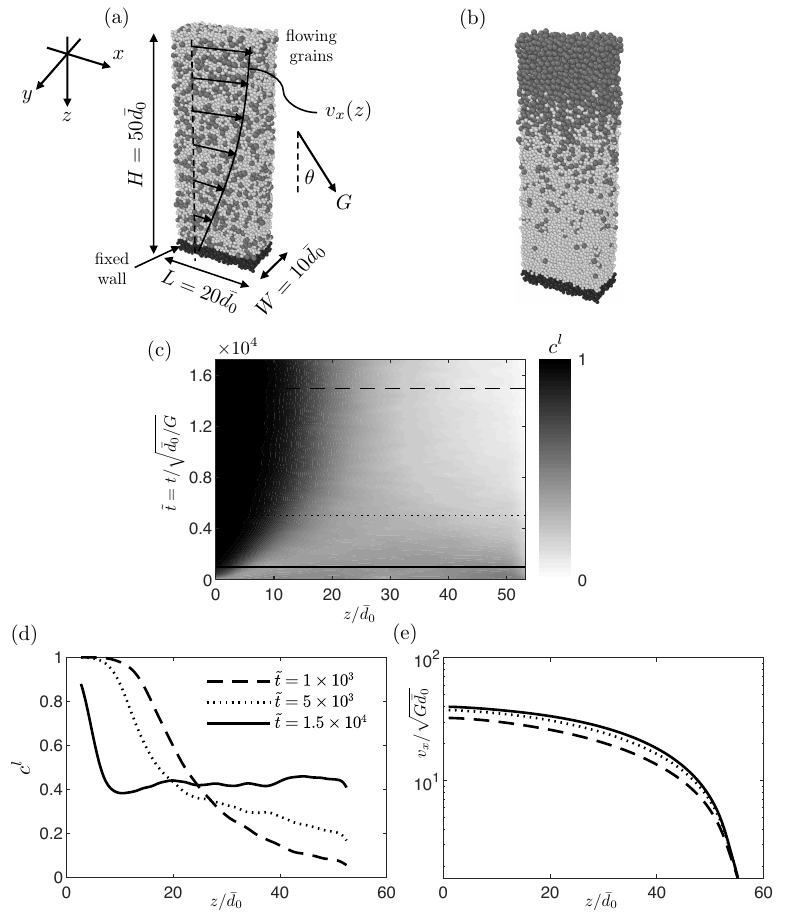}
\end{center}
\caption{(a) Initial well-mixed configuration for three-dimensional DEM simulation of bidisperse inclined plane flow with $\sim$15~000 flowing grains. The thickness of the flowing layer is $H=50 \bar{d}_0$. Black grains represent the rough fixed base. Dark gray grains indicate large flowing grains, and light gray grains indicate small flowing grains. (b) Segregated configuration after flowing for a simulation time of $\tilde{t} = t/\sqrt{\bar{d}_0/G} =  1.5 \times 10^4$. (c) Spatiotemporal evolution of the large-grain concentration field. Spatial profiles of (d) the concentration field $c^{\rm l}$ and (e) the normalized velocity field $v_x/\sqrt{G \bar{d}_0}$ at three times ($\tilde{t} = 1 \times 10^3$, $5 \times 10^3$,  and $1.5 \times 10^4$) as indicated by the horizontal lines in (c).}
\label{fig:IPE_spheres_fig1}
\end{figure}

Consider a semi-infinite layer of thickness $H$ of a dense, bidisperse granular mixture flowing down an inclined plane with surface inclination angle $\theta$ under the action of gravity $G$. For the case of bidisperse spheres, the DEM setup is shown in Fig.~\ref{fig:IPE_spheres_fig1}(a) for $H = 50\bar{d}_0$, where $\bar{d}_0$ is the system-wide average grain size. The large particles are dark gray, and the small particles are light gray. We employ periodic boundary conditions along both the direction of flow (i.e., the $x$-direction) and the lateral direction (i.e., the $y$-direction), eliminating any lateral wall/boundary effects. In all DEM simulations of spheres, we take the length of the simulation domain along the $x$-direction to be $L=20 \bar{d}_0$ and the length along the $y$-direction to be $W=10 \bar{d}_0$, and we have verified that both of these length scales are sufficiently large, so that they do not affect the resulting flow and segregation fields. The rough bottom surface is comprised of touching, glued grains in our DEM simulations, which are denoted as black in Fig.~\ref{fig:IPE_spheres_fig1}(a). This is done to achieve a no-slip boundary condition along the bottom surface. In the resulting flow fields, the only non-zero component of the velocity field is $v_x$, which only varies along the vertical coordinate $z$. A typical steady velocity field is qualitatively sketched in Fig.~\ref{fig:IPE_spheres_fig1}(a), which is consistent with the familiar ``Bagnold profile.''

Regarding the stress field, due to the force balances along the $x$- and $z$-directions, we have that $|\sigma_{xz}(z)| = |\sigma_{zx}(z)| = \phi \rho_{\rm s} G z \sin \theta$ and $\sigma_{zz}(z) = -\phi \rho_{\rm s} G z\cos\theta$, respectively. For two-dimensional systems of disks, we observe in DEM simulations that the normal stresses are approximately equal, i.e., $\sigma_{xx}(z)  \approx \sigma_{zz}(z)$, so that $\tau(z) = | \sigma_{xz}(z) | = \phi \rho_{\rm s} G z \sin \theta$, $P(z) = - \sigma_{zz}(z) = \phi \rho_{\rm s} G z \cos\theta$, and $\mu(z) = \tau(z)/P(z) = \tan\theta$. For three-dimensional systems of spheres, we observe normal stress differences in DEM simulations \citep[e.g.][]{srivastava2021viscometric}, in which, in particular, the magnitude of the out-of-plane normal stress $\sigma_{yy}(z)$ is slightly lower than the magnitude of $\sigma_{zz}(z)$ at each $z$-position. For spheres, the stress ratio $\mu$ is still spatially uniform but slightly higher than $\tan\theta$, and the pressure field $P(z)$ still varies linearly in $z$ but with a slope that is slightly lower than $\phi \rho_{\rm s} G  \cos \theta$.

There are four important dimensionless parameters that specify a case of inclined plane flow: (1) $H/\bar{d}_0$, the dimensionless layer thickness, (2) $\theta$, the inclination angle, which sets the stress field for a given case and controls the total flow rate down the incline, (3) $c^{\rm l}_0(z)$, the initial large-grain concentration, which can be a spatially-varying field; and (4) $d^{\rm l}/d^{\rm s}$, the bidisperse grain size ratio. Thus, the parameter set $\{H/\bar{d}_0, \theta, c^{\rm l}_0, d^{\rm l}/d^{\rm s}\}$ specifies the geometry, loads, and initial conditions for a given case of inclined plane flow. We choose a representative base case for spheres corresponding to the parameter set $\{H/\bar{d}_0=50,\, \theta=26^\circ,\, c^{\rm l}_0 = 0.50, d^{\rm l}/d^{\rm s} = 1.5\}$, and the initially well-mixed DEM configuration for this case is shown in Fig.~\ref{fig:IPE_spheres_fig1}(a). We then run the DEM simulation from this initial configuration and observe the consequent segregation process. Due to the gravitationally-induced pressure gradient, we expect the large grains (dark gray) to be driven towards the top of the layer where the pressure is the lowest.  The segregated state after a simulation time of $\tilde{t} = t/\sqrt{\bar{d}_0/G} = 1.5 \times 10^4$ is shown in Fig.~\ref{fig:IPE_spheres_fig1}(b), demonstrating that the large grains indeed segregate towards the top of the layer, leaving a layer of small grains (light gray) beneath.  In order to obtain a more quantitative picture of the segregation dynamics, we coarse-grain the concentration field $c^{\rm l}$ and plot contours of the spatiotemporal evolution of the $c^{\rm l}$ field in Fig.~\ref{fig:IPE_spheres_fig1}(c). (The methods used for spatial coarse-graining are described in Appendix~\ref{app:coarse_grain}.) The growing width of the dark region indicates the temporal evolution of the large grains segregating to the top of the layer. Further, snapshots of the concentration field $c^{\rm l}$ and the non-dimensionalized velocity field $v_x/\sqrt{G \bar{d}_0}$ at different times---$\tilde{t} = 1 \times 10^3,\, 5 \times 10^3, \,1.5 \times 10^4$  as indicated by the horizontal lines in Fig.~\ref{fig:IPE_spheres_fig1}(c)---are shown in Figs.~\ref{fig:IPE_spheres_fig1}(d) and (e), respectively. It is evident from these plots that the velocity field reaches a steady state early in the simulation time window, while the concentration field continues to evolve and only reaches a steady state towards the end of the simulation time window, i.e., $\tilde{t} \gtrsim 1.5 \times 10^4$. Lastly, the shape of the steady flow field $v_x(z)$ in Fig.~\ref{fig:IPE_spheres_fig1}(e) is consistent the scaling $v_x(z) \propto H^{3/2} - z^{3/2}$, associated with the ``Bagnold profile''  \citep[e.g.,][]{gray05}. 

\begin{figure}[!t]
\begin{center}
\includegraphics{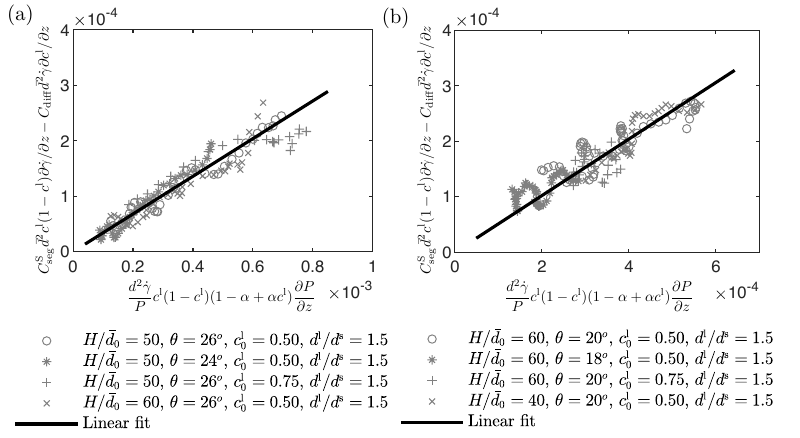}
\end{center}
\caption{Collapse of $C^{\rm S}_{\rm seg} \bar{d}^2c^{\rm l}(1-c^{\rm l})({\partial \dot{\gamma}}/{\partial z}) - C_{\rm diff} \bar{d}^2\dot{\gamma} ({\partial c^{\rm l}}/{\partial z})$  versus $({\bar{d}^2 \dot{\gamma}}/{P}) c^{\rm l}(1-c^{\rm l})(1 - \alpha + \alpha c^{\rm l}) ({\partial P}/{\partial z})$ for several cases of inclined plane flow of (a) bidisperse spheres and (b) bidispserse disks with a size ratio of $d^{\rm l}/d^{\rm s}=1.5$. Symbols represent coarse-grained, steady-state DEM field data, and the solid lines are the best linear fits using (a) $C^{\rm P}_{\rm seg} = 0.34$ for spheres and (b) $C^{\rm P}_{\rm seg} = 0.51$ for disks.}
\label{fig:flux_balance}
\end{figure}

In the inclined plane flow configuration, both pressure gradients and shear-strain-rate gradients are present. The shape of the steady flow field $v_x(z)$ in Fig.~\ref{fig:IPE_spheres_fig1}(d) indicates that the shear strain rate is greatest at the bottom of the layer, decreasing with the distance from the bottom of the layer and approaching zero at the top of the layer. Therefore, the shear-strain-rate gradient drives the large grains towards the high-strain-rate region at the bottom of the layer, while the pressure gradient drives the large grains towards the low-pressure region at the top of the layer. As a result, the shear-strain-rate-gradient-driven and pressure-gradient-driven fluxes are in competition with one another, and the pressure-gradient-driven flux clearly wins out and drives the large grains towards the top of the layer. This characteristic of inclined plane flow makes it a suitable configuration for estimating the dimensionless material parameters $\{C^{\rm P}_{\rm seg},\alpha\}$. To achieve this, we first run the DEM simulation long enough so that all evolving fields reach the steady-state regime ($\tilde{t} \gtrsim 1.5 \times 10^4$). In this regime, ${D c^{\rm l}}/{D t} \approx 0 $, and according to the balance of mass equation \eqref{eq:cons_mass} and the no-flux boundary conditions at the top and bottom of the layer, the total flux is zero, i.e., $w^{\rm l}_z = w^{\rm diff}_z + w^{\rm S}_z + w^{\rm P}_z  = 0$, at each $z$-position. Therefore, in the steady-state regime, using \eqref{eq:diff_eqn}, \eqref{eq:segS_eqn}, and \eqref{eq:segP_eqn}, we have that 
\begin{equation}\label{eq:flux_balance}
C^{\rm S}_{\rm seg}\bar{d}^2c^{\rm l}(1-c^{\rm l})\frac{\partial \dot{\gamma}}{\partial z} - C_{\rm diff} \bar{d}^2\dot{\gamma} \frac{\partial c^{\rm l}}{\partial z} = C^{\rm P}_{\rm seg} \frac{\bar{d}^2 \dot{\gamma}}{P} c^{\rm l}(1-c^{\rm l})(1 - \alpha + \alpha c^{\rm l}) \frac{\partial P}{\partial z}.
\end{equation}
Since the parameters $C_{\rm diff}$ and $C^{\rm S}_{\rm seg}$ have been previously determined, we use this steady-state flux balance to estimate the parameters $\{C^{\rm P}_{\rm seg},\alpha\}$. To do so, we spatially coarse-grain the DEM data from $1000$ evenly-distributed snapshots in the steady-state regime to obtain the relevant field quantities in \eqref{eq:flux_balance}---namely, $c^{\rm l}$ (and hence $\bar{d}$), $\partial c^{\rm l}/\partial z$, $v_x$, $\dot{\gamma} = \partial v_x/\partial z$, $\partial \dot{\gamma}/\partial z = \partial^2 v_x/ \partial z^2$, $P$, and $\partial P/\partial z$. The fields are then arithmetically averaged in time so that the resultant fields only depend on the spatial coordinate $z$. Equation~\eqref{eq:flux_balance} suggests determining the parameter $C^{\rm P}_{\rm seg}$ from the slope of the $C^{\rm S}_{\rm seg} \bar{d}^2c^{\rm l}(1-c^{\rm l})({\partial \dot{\gamma}}/{\partial z}) - C_{\rm diff} \bar{d}^2\dot{\gamma} ({\partial c^{\rm l}}/{\partial z})$  versus $({\bar{d}^2 \dot{\gamma}}/{P}) c^{\rm l}(1-c^{\rm l})(1 - \alpha + \alpha c^{\rm l}) ({\partial P}/{\partial z})$ relation for a given choice of $\alpha \, \in \,[0,1]$. Since $\alpha$ is an adjustable parameter along with $C^{\rm P}_{\rm seg}$, additional data is helpful to more precisely estimate the values of the parameters $\{C^{\rm P}_{\rm seg},\alpha\}$, and we consider three additional variants of the base case---namely, (1) a lower inclination angle case $\{H/\bar{d}_0=50,\, \theta=24^\circ,\, c^{\rm l}_0 = 0.50, d^{\rm l}/d^{\rm s} = 1.5\}$, (2) a more large grains case $\{H/\bar{d}_0=50,\, \theta=26^\circ,\, c^{\rm l}_0 = 0.75, d^{\rm l}/d^{\rm s} = 1.5\}$, and (3) a thicker layer case $\{H/\bar{d}_0=60,\, \theta=26^\circ,\, c^{\rm l}_0 = 0.50, d^{\rm l}/d^{\rm s} = 1.5\}$. We coarse-grain the steady-state DEM field data for all four cases and iterate over values of $\alpha$, seeking the strongest linear collapse in $C^{\rm S}_{\rm seg} \bar{d}^2c^{\rm l}(1-c^{\rm l})({\partial \dot{\gamma}}/{\partial z}) - C_{\rm diff} \bar{d}^2\dot{\gamma} ({\partial c^{\rm l}}/{\partial z})$  versus $({\bar{d}^2 \dot{\gamma}}/{P}) c^{\rm l}(1-c^{\rm l})(1 - \alpha + \alpha c^{\rm l}) ({\partial P}/{\partial z})$. The DEM data for the best-fit case of $\alpha=0.4$ is shown in Fig.~\ref{fig:flux_balance}(a), where each data point represents a unique $z$-position. The data collapses quite well across the different cases, and a linear dependence is evident. For $\alpha=0.4$, the coefficient of determination (i.e., the R-squared value) is $0.92$. Finally, the dimensionless material parameter $C^{\rm P}_{\rm seg}$ may be obtained from the slope of the linear relation in Fig.~\ref{fig:flux_balance}(a) (indicated by the solid line), which we determine to be $C^{\rm P}_{\rm seg} = 0.34$ for frictional spheres with a size ratio of $d^{\rm l}/d^{\rm s}=1.5$. 

We also apply this process to dense, bidisperse mixtures of frictional disks to determine the parameters $\{C^{\rm P}_{\rm seg},\alpha\}$. We consider a base case for disks corresponding to the parameter set $\{H/\bar{d}_0=60,\, \theta=20^\circ,\, c^{\rm l}_0 = 0.50, d^{\rm l}/d^{\rm s} = 1.5\}$ as well as three variants---(1) a lower inclination angle case $\{H/\bar{d}_0=60,\, \theta=18^\circ,\, c^{\rm l}_0 = 0.50, d^{\rm l}/d^{\rm s} = 1.5\}$, (2) a more large grains case $\{H/\bar{d}_0=60,\, \theta=20^\circ,\, c^{\rm l}_0 = 0.75, d^{\rm l}/d^{\rm s} = 1.5\}$, and (3) a thinner layer case $\{H/\bar{d}_0=40,\, \theta=20^\circ,\, c^{\rm l}_0 = 0.50, d^{\rm l}/d^{\rm s} = 1.5\}$. In the DEM simulations for disks, the length of the simulation domain along the $x$-direction is taken to be $L=60\bar{d}_0$ in all cases. Each DEM simulation is run to steady-state; the steady-state DEM data is coarse-grained; and the flux balance \eqref{eq:flux_balance} is applied. The DEM data for all four cases is plotted in Fig.~\ref{fig:flux_balance}(b) for $\alpha=0.4$, and the data again collapses quite well. A linear relation is observed (with an R-squared value of $0.82$), and the parameter $C^{\rm P}_{\rm seg}$ for frictional disks with a size ratio of $d^{\rm l}/d^{\rm s}=1.5$ is determined from the slope of the linear relation to be $C^{\rm P}_{\rm seg} = 0.51$. We note that we have utilized the same value of $\alpha=0.4$ for both spheres and disks. Of course, the parameter $\alpha$ may be fitted separately to steady-state DEM data for spheres and disks, respectively, but we observed that the separately fitted values of $\alpha$ turn out to be quite similar, so for simplicity, the value of $\alpha=0.4$ represents the collective best fit that yields the strongest linear collapses of the steady-state DEM data for both spheres and disks simultaneously. 

\section{Validation of the continuum model in the transient regime} \label{sec:validation} 
In Section~\ref{sec:seg_model}, we extended the coupled model for segregation and flow in the absence of pressure gradients proposed in our prior work \citep{liu_singh_arxiv} to account for pressure-gradient-driven segregation, and in Section~\ref{sec:PGDS}, we estimated the dimensionless material parameters $\{C^{\rm P}_{\rm seg},\alpha\}$ associated with the pressure-gradient-driven segregation flux using steady-state DEM data in inclined plane flow. In this section, we test the continuum model by comparing predictions of the transient evolution of the large-grain concentration fields and the steady-state flow fields against DEM simulation results. To this end, we consider both inclined plane flow as described in Section~\ref{sec:PGDS} and an additional flow geometry---planar shear flow with gravity. The coupled continuum model consists of the segregation dynamics equation \eqref{eq:segP_model_seg} and the NGF model, \eqref{eq:flow_rule} and \eqref{eq:nl_rheology}, and throughout, we utilize a fixed set of parameters for spheres, 
\begin{equation}\label{mater_param_seg_spheres}
\{ \mu_{\rm s}=0.37, \mu_2=0.95, I_0=0.58, A=0.43, C_{\rm diff}=0.045, C^{\rm S}_{\rm seg}=0.08, C^{\rm P}_{\rm seg}=0.34, \alpha=0.4 \},
\end{equation}
and for disks,
\begin{equation}\label{mater_param_seg_disks}
\{ \mu_{\rm s}=0.272, b=1.168, A=0.90, C_{\rm diff}=0.20, C^{\rm S}_{\rm seg}=0.23, C^{\rm P}_{\rm seg}=0.51, \alpha=0.4 \}.
\end{equation}

\subsection{Inclined plane flow}\label{sec:IPF_transient_comparisons}
We first consider inclined plane flow to test predictions of the coupled continuum model against corresponding DEM results. As discussed above, when there are no normal stress differences, the stress field may be obtained from a static force balance, which implies that the stress ratio field is uniform and given by $\mu(z) = \tan \theta$ and that the pressure field is linear in $z$ and given by $P(z) = \phi \rho_{\rm s} G z \cos\theta$. However, the normal stress differences that arise in dense flows of spheres induce a slightly higher, uniform value of $\mu$ and a slightly lower slope in the pressure field $P(z)$. Accordingly, to control for this effect when working with dense flows of spheres, in our continuum simulations, we utilize the values of the uniform stress ratio and the slope of the pressure field obtained from the coarse-grained stress fields in the DEM data for each case, rather than the nominal values of $\tan\theta$ and $\phi \rho_{\rm s} G \cos\theta$, respectively. Moreover, to avoid a singularity in the pressure-gradient-driven segregation flux \eqref{eq:segP_eqn} at the free surface where $z=0$, we add a small constant to the pressure field corresponding to the weight of a layer of $(1/4)\bar{d}_0$ thickness, i.e., $(1/4)\phi \rho_{\rm s} G \bar{d}_0 \cos\theta$. This approach is quite similar to that of \citet{trewhela2021experimental}, who directly incorporate this small constant into the denominator of their pressure-gradient-driven flux equation. We have verified that the subsequently presented results are insensitive to the exact choice of this constant, so long as it is sufficiently small. With the relevant stress-related fields determined in this way, the balance of linear momentum \eqref{eq:EOM} is satisfied and does not further enter the solution procedure. 

Continuum model predictions are obtained by numerically solving the remaining governing equations using finite-differences. The remaining unknown fields in inclined plane flow are the velocity field $v_x(z,t)$, the strain-rate field $\dot{\gamma}(z,t) = \partial v_x/\partial z$, the granular fluidity field $g(z,t)$, and the concentration field $c^{\rm l}(z,t)$. The governing equations are (1) the flow rule \eqref{eq:flow_rule}, 
\begin{equation}\label{eq:flow_rule2}
\dot{\gamma} = g \mu,
\end{equation}
(2) the nonlocal rheology \eqref{eq:nl_rheology}, 
\begin{equation}\label{eq:nl_rheology2}
g = g_{\rm loc} (\mu, P) + \xi^2(\mu) \frac{\partial^2 g}{\partial z^2},
\end{equation}
where $g_{\rm loc} (\mu, P)$ and $\xi(\mu)$ are given through \eqref{eq:localg_bi_seg} and \eqref{eq: cooperativity_bi_seg}$_1$, respectively, and finally (3) the segregation dynamics equation \eqref{eq:segP_model_seg},
\begin{equation}\label{eq:conc_evol}
\dfrac{\partial c^{\rm l}}{\partial t} + \frac{\partial}{\partial z}  \left(- C_{\rm diff} \bar{d}^2\dot{\gamma} \frac{\partial c^{\rm l}}{\partial z} + C^{\rm S}_{\rm seg}\bar{d}^2c^{\rm l}(1-c^{\rm l})\frac{\partial \dot{\gamma}}{\partial z} - C^{\rm P}_{\rm seg} \frac{\bar{d}^2 \dot{\gamma}}{P} c^{\rm l}(1-c^{\rm l})(1 - \alpha + \alpha c^{\rm l}) \frac{\partial P}{\partial z} \right) = 0,
\end{equation}
where $\bar{d} = c^{\rm l} d^{\rm l} + (1-c^{\rm l}) d^{\rm s}$. 

For the fluidity boundary conditions, we impose a Neumann boundary condition for the fluidity at the free surface, i.e., $\partial g/ \partial z =0$ at $z=0$, and a Dirichlet fluidity boundary condition at the bottom surface, i.e., $g=g_{\rm loc}(\mu(z=H), P(z=H))$ at $z=H$. For the segregation dynamics equation \eqref{eq:conc_evol}, we apply no flux boundary conditions at both the free surface and the bottom surface, i.e., $w^{\rm l}_z = - C_{\rm diff} \bar{d}^2 \dot{\gamma} (\partial c^{\rm l}/ \partial z ) + C^{\rm S}_{\rm seg} \bar{d}^2 c^{\rm l} (1- c^{\rm l})( \partial \dot{\gamma} / \partial z ) - C^{\rm P}_{\rm seg} ({\bar{d}^2 \dot{\gamma}}/{P})c^{\rm l}(1- c^{\rm l})(1 - \alpha + \alpha c^{\rm l}) (\partial P/ \partial z) =0 $ at $z=0$ and $H$. Finally, an initial condition for the segregation dynamics equation \eqref{eq:conc_evol}---i.e., $c^{\rm l}_0(z) = c^{\rm l}(z, t=0)$---is necessary. To account for spatial fluctuations in $c^{\rm l}_0(z)$, we coarse-grain the concentration field in the DEM configuration at $t=0$ for each case and use this field as the initial condition in the corresponding continuum simulation.

Then, we obtain numerical predictions of the continuum model for a given case of inclined plane flow using finite differences as follows. At a given point in time, the concentration field $c^{\rm l}(z)$ is known, and thus, the average grain-size may be calculated as $\bar{d}(z) = c^{\rm l}(z)d^{\rm l} + (1-c^{\rm l}(z))d^{\rm s}$. Since the pressure distribution $P(z)$ and stress ratio distribution $\mu(z)$ are known, the local fluidity $g_{\rm loc}(\mu, P)$ and the cooperativity length $\xi(\mu)$, given through equations \eqref{eq:localg_bi_seg} and \eqref{eq: cooperativity_bi_seg}$_1$, respectively, can also be calculated at the spatial grid points. The nonlocal rheology equation \eqref{eq:nl_rheology2} can then be solved by discretizing the Laplacian term $\partial^2 g / \partial z^2$ using central differences in space. Once the fluidity field has been calculated at all grid points, the strain-rate field can be calculated using equation \eqref{eq:flow_rule2}, and the velocity field $v_x(z)$ can be obtained by integrating the strain-rate field. Next, the segregation dynamics equation \eqref{eq:conc_evol} is used to update the concentration by (1) using a forward Euler discretization for $\partial {c}^{\rm l}/\partial t$, (2) treating both of the segregation fluxes explicitly, and (3) treating the diffusion term implicitly. We have verified that both the spatial and temporal resolutions employed are sufficiently refined, so as to ensure the numerical accuracy and stability of our finite-difference scheme. With this, one step of numerical integration is completed and the unknown fields $c^{\rm l}$, hence $\bar{d}$, are obtained at the next time step. The same procedure is repeated over multiple time steps to calculate the transient evolution of the concentration and flow fields, $c^{\rm l}(z,t)$ and $v_x(z,t)$, over the desired simulation time window.

\begin{figure}
\begin{center}
\includegraphics[width=0.8\columnwidth]{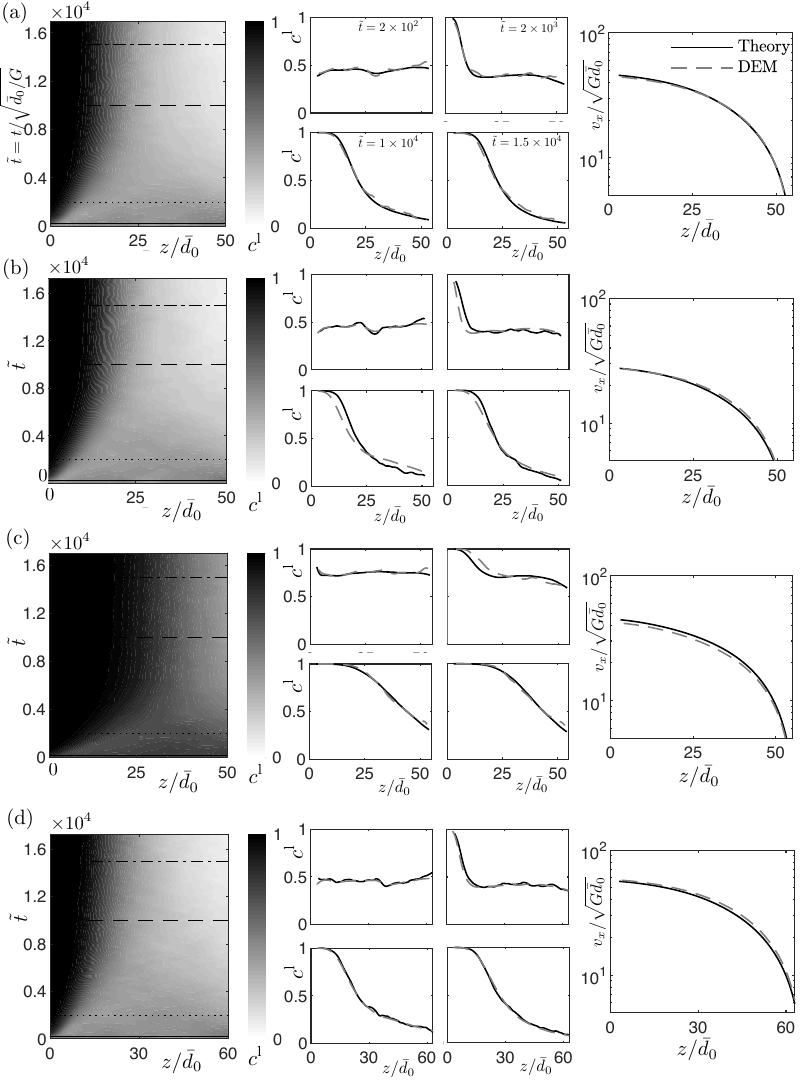}
\end{center}
\caption{Comparisons of continuum model predictions with corresponding DEM simulation results for the transient evolution of the segregation dynamics for four cases of inclined plane flow of bidisperse spheres: (a) Base case $\{H/\bar{d}_0=50,\, \theta=26^\circ,\, c^{\rm l}_0 = 0.50, d^{\rm l}/d^{\rm s} = 1.5\}$; (b) Lower inclination angle case $\{H/\bar{d}_0=50,\, \theta=24^\circ,\, c^{\rm l}_0 = 0.50, d^{\rm l}/d^{\rm s} = 1.5\}$; (c) More large grains case $\{H/\bar{d}_0=50,\, \theta=26^\circ,\, c^{\rm l}_0 = 0.75, d^{\rm l}/d^{\rm s} = 1.5\}$; and (d) Thicker layer case $\{H/\bar{d}_0=60,\, \theta=26^\circ,\, c^{\rm l}_0 = 0.50, d^{\rm l}/d^{\rm s} = 1.5\}$. For each case, the first column shows spatiotemporal contours of the evolution of $c^{\rm l}$  measured in DEM simulations. The second column shows comparisons of the DEM simulations (dashed gray lines) and continuum model predictions (solid black lines) of the $c^{\rm l}$ field at four different time instants during the segregation process: $\tilde{t}  = 2 \times 10^2$, $2 \times 10^3$, $1 \times 10^4$, and $1.5 \times 10^4 $ in the sequence of top left, top right, bottom left, bottom right. The third column shows comparisons of the steady-state, normalized velocity profiles at $\tilde{t} = 1.5 \times 10^4$ from DEM simulations and continuum model predictions.}
\label{fig:SegP_IPF_transient_comparisons_spheres}
\end{figure}

We compare model predictions of the transient evolution of the concentration field and the steady-state flow field against corresponding DEM simulation results for all four cases of inclined plane flow of dense, bidisperse spheres considered in Section~\ref{sec:PGDS}. The comparisons are summarized in Fig.~\ref{fig:SegP_IPF_transient_comparisons_spheres}. For each case, the spatiotemporal evolution of the DEM-measured $c^{\rm l}$ field is shown in the first column of Fig.~\ref{fig:SegP_IPF_transient_comparisons_spheres}, and the second column shows comparisons of  snapshots of the $c^{\rm l}$ field measured in DEM simulations against corresponding continuum model predictions at four different time instants---$\tilde{t} =  t/\sqrt{\bar{d}_0/G} =  2 \times 10^2,\, 2 \times 10^3,\,1 \times 10^4$, and $1.5 \times 10^4 $---as indicated by the horizontal lines on the contour plots in the first column. The model does a good job predicting the segregation dynamics across different cases of inclined plane flow and can capture the variations in the evolution of the $c^{\rm l}$ field as the input parameters are changed. The third column shows comparisons of the steady-state velocity field predicted by the continuum model with the corresponding DEM-measured velocity fields. The Bagnold-like profile is captured well in all cases. We note that in inclined plane flow, the local inertial, or $\mu(I)$, rheology described in Appendix~\ref{app:local} would be sufficient to predict the flow fields, and a nonlocal rheological approach is not necessary for these flows. However, as shown in our prior work \citep{liu_singh_arxiv} and in the subsequent section on planar shear flow with gravity, nonlocal rheological modeling is crucial in many other flow configurations. One benefit of the NGF model is that it automatically reduces to the local inertial rheology when appropriate, so for consistency we have continued to use the NGF model when considering inclined plane flow.

\begin{figure}[!t]
\begin{center}
\includegraphics[width=0.8\columnwidth]{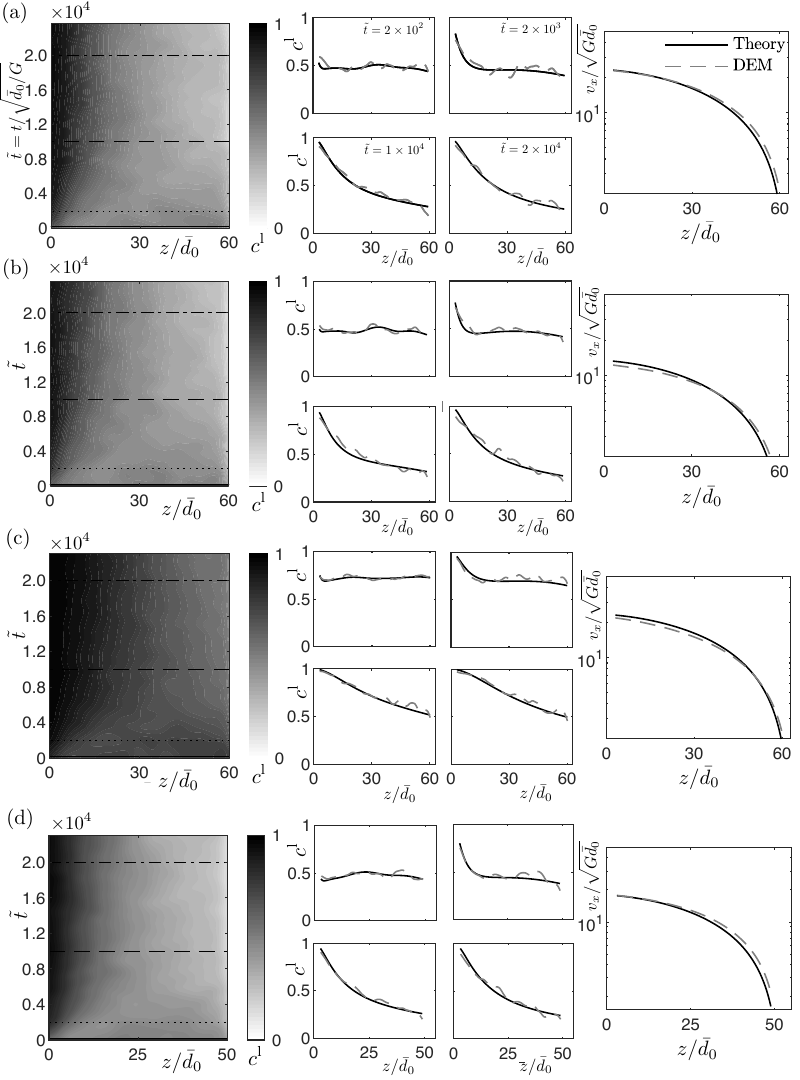}
\end{center}
\caption{Comparisons of continuum model predictions with corresponding DEM simulation results for the transient evolution of the segregation dynamics for four cases of inclined plane flow of bidisperse disks: (a) Base case $\{H/\bar{d}_0=60,\, \theta=20^\circ,\, c^{\rm l}_0 = 0.50, d^{\rm l}/d^{\rm s} = 1.5\}$; (b) Lower inclination angle case $\{H/\bar{d}_0=60,\, \theta=18^\circ,\, c^{\rm l}_0 = 0.50, d^{\rm l}/d^{\rm s} = 1.5\}$; (c) More large grains case $\{H/\bar{d}_0=60,\, \theta=20^\circ,\, c^{\rm l}_0 = 0.75, d^{\rm l}/d^{\rm s} = 1.5\}$; and (d) Thinner layer case $\{H/\bar{d}_0=50,\, \theta=26^\circ,\, c^{\rm l}_0 = 0.50, d^{\rm l}/d^{\rm s} = 1.5\}$. Results are organized as described in the caption of Fig.~\ref{fig:SegP_IPF_transient_comparisons_spheres}. The time instants shown in the second column correspond to $\tilde{t} = 2 \times 10^2, \, 2 \times 10^3, \, 1 \times 10^4$, and  $2 \times 10^4$.}
\label{fig:SegP_IPF_transient_comparisons_disks}
\end{figure}

We also test predictions of the coupled continuum model against corresponding DEM results for dense, bidisperse flows of disks. The governing equations and boundary conditions are the same as those used above for dense, bidisperse flows of spheres but with $g_{\rm loc}(\mu,P)$ and $\xi(\mu)$ given through \eqref{eq:localg_bi_seg_disks} and \eqref{eq: cooperativity_bi_seg}$_2$, respectively, and using the material parameter set for disks given in \eqref{mater_param_seg_disks}. One modification is that for disks, we use the stress ratio and pressure fields based on a static force balance---i.e., $\mu(z)=\tan\theta$ and $P(z) =\phi\rho_{\rm s} Gz\cos\theta$ with the same small constant added to the pressure field---instead of using fields based on coarse-grained DEM results. This is because normal stress differences are minimal in two-dimensional flows of disks, and the DEM-measured stress fields are quite close to the corresponding fields derived based on force balances. To test the continuum model, we consider four cases of inclined plane flow of disks: (1) the base case $\{H/\bar{d}_0=60,\, \theta=20^\circ,\, c^{\rm l}_0 = 0.50, d^{\rm l}/d^{\rm s} = 1.5\}$, (2) a lower inclination angle case $\{H/\bar{d}_0=60,\, \theta=18^\circ,\, c^{\rm l}_0 = 0.50, d^{\rm l}/d^{\rm s} = 1.5\}$, (3) a more large grains case $\{H/\bar{d}_0=60,\, \theta=20^\circ,\, c^{\rm l}_0 = 0.75, d^{\rm l}/d^{\rm s} = 1.5\}$, and (4) a thinner layer case $\{H/\bar{d}_0=50,\, \theta=20^\circ,\, c^{\rm l}_0 = 0.50, d^{\rm l}/d^{\rm s} = 1.5\}$, and results are shown in Figs.~\ref{fig:SegP_IPF_transient_comparisons_disks} (a), (b), (c), and (d), respectively. The first column shows contour plots of the spatiotemporal evolution of the $c^{\rm l}$ field obtained from DEM simulations, and the second column compares continuum model predictions of the $c^{\rm l}$ field against DEM simulations at four different time instants---$\tilde{t} = 2 \times 10^2, \, 2 \times 10^3, \, 1 \times 10^4$, and  $2 \times 10^4$---as indicated by the horizontal lines on the contour plot in the first column for each case. The last column compares the steady-state flow field predicted by the continuum model against the DEM-measured flow fields.  The model again does a good job simultaneously predicting the segregation dynamics and the steady-state flow fields in dense, bidisperse flows of frictional disks. 

\subsection{Planar shear flow with gravity} \label{sec:PSwG_transient_comparisons} 
We have hypothesized that the proposed constitutive equation \eqref{eq:segP_eqn} for the pressure-gradient-driven segregation flux and the dimensionless material parameters $\{C^{\rm P}_{\rm seg},\alpha\}$ are general across different flow geometries. To test this hypothesis, we consider a new flow configuration: planar shear flow with gravity acting orthogonal to the shearing direction. We did not use DEM data from flows in this configuration to estimate the parameters $\{C^{\rm P}_{\rm seg},\alpha\}$ and do no further parameter adjustment in this section, so comparisons of continuum model predictions of the transient evolution of the large-grain concentration field and the steady-state flow field against DEM simulation results may be regarded as independent validation tests of the model.

\begin{figure}[!t]
\begin{center}
\includegraphics{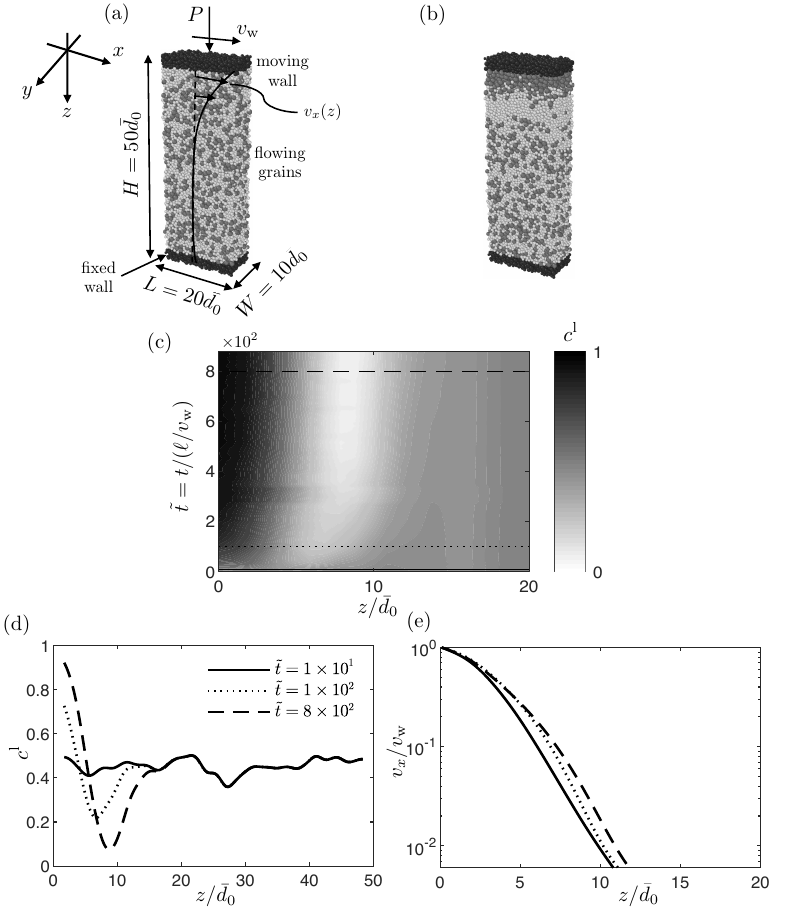}
\end{center}
\caption{Representative base case for planar shear flow with gravity. (a) Initial well-mixed configuration for three-dimensional DEM simulation of a bidisperse mixture of spheres with $\sim$15~000 flowing grains. The thickness of the flowing layer is $H=50 \bar{d}_0$. Black grains represent the rough top and bottom walls. Dark gray grains indicate large flowing grains, and light gray grains indicate small flowing grains. (b) Segregated state after flowing for a simulation time of $\tilde{t} = t/(\ell/v_{\rm w}) =  8 \times 10^2$.  (c) Spatiotemporal evolution of the large-grain concentration field. Spatial profiles of (d) the concentration field $c^{\rm l}$ and (e) the normalized velocity field $v_x/v_{\rm w}$ at three times ($\tilde{t} = 1 \times 10^1, 1 \times 10^2$, and $8 \times 10^2$) as indicated by the horizontal lines in (c).}
\label{fig:PSwG_fig1}
\end{figure}

We first describe this flow configuration in detail and discuss its important characteristics. Consider a semi-infinite layer of a dense, bidisperse granular mixture between two rough parallel walls, separated by a distance $H$ along the $z$-direction. For the case of bidisperse spheres, the DEM setup is shown in Fig.~\ref{fig:PSwG_fig1}(a) for $H=50 \bar{d}_0$. The top and bottom walls, indicated as black grains in Fig.~\ref{fig:PSwG_fig1}(a), are made of layers of touching, glued grains. The top wall is subjected to a compressive normal stress $P_{\rm w}$ along the $z$-direction and is specified to move with a constant velocity $v_{\rm w}$ along the $x$-direction. The bottom wall is specified to remain fixed and imposes a no-slip boundary condition at the bottom. The gravitational body force with acceleration of gravity $G$ acts along the $z$-direction.  Furthermore, we employ periodic boundary conditions along the flow ($x$) and lateral ($y$) directions, eliminating any lateral wall/boundary effects and leading to a simple one-dimensional flow in which the only non-zero velocity field component is $v_x$ that depends only on the $z$-coordinate. In all DEM simulations of spheres, we take the length of the simulation domain along the $x$-direction to be $L = 20 \bar{d}_0$ and the length along the $y$-direction to be $W = 10 \bar{d}_0$, as shown in Fig.~\ref{fig:PSwG_fig1}(a). 

Regarding the stress field, the force balance along the $x$-direction gives that the shear stress magnitude is spatially uniform and equal to the shear stress imparted by the moving wall $\tau_{\rm w}$, i.e., $|\sigma_{xz}(z)| = |\sigma_{zx}(z)| =  \tau_{\rm w}$. We note that instead of directly prescribing $\tau_{\rm w}$ in our DEM simulations, the top-wall velocity $v_{\rm w}$ is prescribed, and the shear stress $\tau_{\rm w}$ arises as an output. The force balance along the $z$-direction gives that $\sigma_{zz}(z) = -(P_{\rm w} + \phi \rho_{\rm s} G z)$. As discussed in Section~\ref{sec:PGDS}, we observe that $\sigma_{xx}(z) \approx \sigma_{zz}(z)$ in DEM simulations of two-dimensional systems of disks, so that $\tau(z) = |\sigma_{xz}(z)| = \tau_{\rm w}$, $P(z) = -\sigma_{zz} = P_{\rm w} + \phi \rho_{\rm s} G z$, and $\mu(z) = \tau(z)/P(z) = {\mu_{\rm w}}/{(1 + z/\ell)}$, where $\mu_{\rm w} = \tau_{\rm w}/P_{\rm w}$ is the maximum value of the stress ratio occurring at the top wall ($z=0$) and $\ell=P_{\rm w}/\phi \rho_{\rm s} G$ is the loading length scale. The loading length scale is defined as the ratio of the top-wall pressure to the gravitational body force and is a distinct length scale from the dimensions $H$, $L$, and $W$. Again, while the coarse-grained stress fields in DEM simulations of disks are quite close to these analytical fields determined from force balances, in DEM simulations of spheres, we observe normal stress differences, and while this leads to slight differences in the coarse-grained $\mu(z)$ and $P(z)$ fields for spheres, the $z$-dependence of these fields is consistent with the expressions derived above. Since the stress ratio is greatest at the top wall ($z=0$) and decays with $z$, we expect there to be a flowing region just beneath the top wall with the velocity field $v_x(z)$ decaying as $z$ increases, as qualitatively sketched in Fig.~\ref{fig:PSwG_fig1}(a). Importantly, the loading length-scale $\ell$ is the only length scale appearing in the analytical expression for the stress ratio field  $\mu(z) = {\mu_{\rm w}}/{(1 + z/\ell)}$. Therefore, $\ell$ and not the layer thickness $H$ is the relevant length scale in planar shear flow with gravity and affects the characteristic size of the flowing region beneath the top wall.  We have verified that $H=50 \bar{d}_0$ is sufficiently large, so that it does not affect the flow fields or the segregation dynamics.

The important dimensionless parameters that describe planar shear flow with gravity are (1) ${\ell}/{\bar{d}_0}$, the dimensionless loading length scale, (2) $\tilde{v}_{\rm w}=({v_{\rm w}}/{\ell}) \sqrt{{\rho_{\rm s} \bar{d}_0^2}/{P_{\rm w}}}$, the dimensionless top-wall velocity that determines shear stress  $\tau_{\rm w}$ and the stress ratio $\mu_{\rm w}$ at the top wall, (3) $c^{\rm l}_0(z)$, the initial large grain concentration field, and (4) $d^{\rm l}/d^{\rm s}$, the bidisperse grain-size ratio. Thus, the parameter set $\{\ell/\bar{d}_0, \tilde{v}_{\rm w}, c^{\rm l}_0, d^{\rm l}/d^{\rm s}\}$ fully specifies a given case of planar shear flow with gravity. We consider a representative base case for spheres corresponding to the parameter set $\{\ell/\bar{d}_0=18,\,\tilde{v}_{\rm w}=0.02, \, c^{\rm l}_0=0.50, \, d^{\rm l}/d^{\rm s} = 1.5\}$. In this flow configuration, the strain rate is greatest just under the top wall, where the pressure is lowest. Therefore, the shear-strain-rate-gradient-driven flux and the pressure-gradient-driven flux cooperate and drive large grains towards the top of the layer. The segregated state after running the base-case DEM simulation to a time of  $\tilde{t} = t/(\ell/v_{\rm w}) = 8 \times 10^2$ is shown in Fig.~\ref{fig:PSwG_fig1}(b), illustrating that the large grains (dark gray) indeed segregate towards the top wall, leaving a small-grain-rich region (light gray) underneath. Further away from the wall, the evolution of segregation is very slow due to the low strain rates in that region, and the bidisperse granular material remains well-mixed. For a more quantitative picture, we coarse-grain the concentration field $c^{\rm l}$ in both space and time, and the spatiotemporal evolution of $c^{\rm l}$ is shown in the contour plot in Fig.~\ref{fig:PSwG_fig1}(c). Lastly, snapshots of the $c^{\rm l}$ field and the non-dimensionalized velocity field $v_x/v_{\rm w}$ at three different times---$\tilde{t} = 1 \times 10^1,\, 1 \times 10^2$, and $8 \times 10^2$---are shown in Figs.~\ref{fig:PSwG_fig1}(d) and (e), respectively. Again, we observe that the velocity field quickly reaches a steady state, while the concentration field evolves over the full simulation time window. 

Next, we solve the coupled continuum model for the transient evolution of the concentration and velocity fields in this geometry, following a process analogous to that described in Section~\ref{sec:IPF_transient_comparisons} for inclined plane flow. To control for the effects of normal stress differences in dense flows of spheres, the pressure at the top wall $P(z=0)$ and the slope of the pressure field are obtained from the coarse-grained pressure field in the DEM data for each case. These values are slightly different than the nominal values of $P_{\rm w}$ and $\phi\rho_{\rm s}G$ and give rise to a slightly adjusted loading length scale $\ell$, and we utilize these adjusted values in the stress fields, $\mu(z)$ and $P(z)$, in our continuum simulations. The governing equations are the flow rule \eqref{eq:flow_rule2}, the nonlocal rheology \eqref{eq:nl_rheology2}, and the segregation dynamics equation \eqref{eq:conc_evol}, wherein $\mu(z)$ and $P(z)$ are given as inputs to the model. For the boundary conditions, we impose Dirichlet fluidity boundary conditions at both the top and bottom walls, i.e., $g=g_{\rm loc}(\mu(z=0),P(z=0))$ at $z=0$ and $g=g_{\rm loc}(\mu(z=H),P(z=H))$ at $z=H$, and no-flux boundary conditions at the top and bottom walls, i.e., $w^{\rm l}_z=0$ at $z=0$ and $z=H$. The initial concentration field $c^{\rm l}_0(z)$ is extracted from the initial DEM configuration for each case, and we use the same finite-difference-based numerical approach described above for inclined plane flow. Finally, since $v_{\rm w}$ is prescribed in the DEM simulations of planar shear flow with gravity, while $\mu_{\rm w}$ is specified in the corresponding continuum simulations, we iteratively adjust the value of $\mu_{\rm w}$ input into the continuum simulations to match the target value of $v_{\rm w}$ in the predicted steady-state velocity field.

\begin{figure}
\begin{center}
\includegraphics[width=0.79\columnwidth]{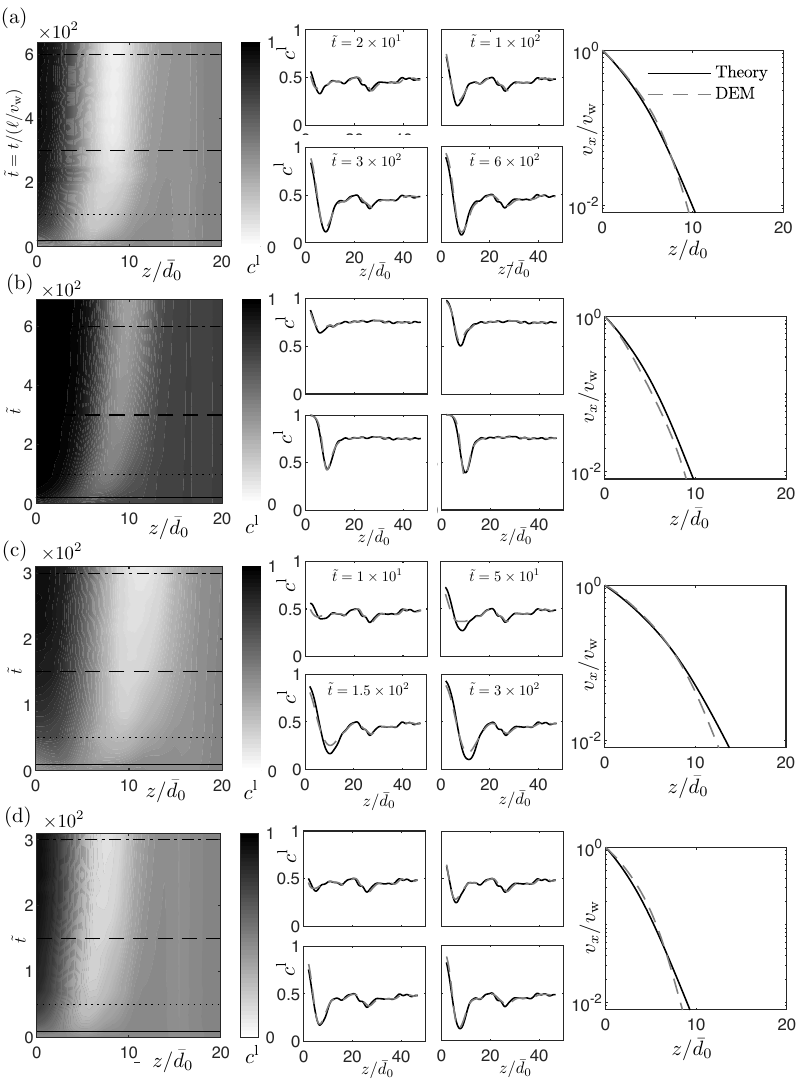}
\end{center}
\caption{Comparisons of continuum model predictions with corresponding DEM simulation results for the transient evolution of the segregation dynamics for four cases of planar shear flow with gravity of bidisperse spheres: (a) Base case $\{\ell/\bar{d}_0=18,\,\tilde{v}_{\rm w}=0.02, \, c^{\rm l}_0=0.50, \, d^{\rm l}/d^{\rm s} = 1.5\}$; (b) More large grains case $\{\ell/\bar{d}_0=18,\,\tilde{v}_{\rm w}=0.02, \, c^{\rm l}_0=0.75, \, d^{\rm l}/d^{\rm s} = 1.5\}$; (c) Larger loading length scale case $\{\ell/\bar{d}_0=36,\,\tilde{v}_{\rm w}=0.01 \, c^{\rm l}_0=0.50, \, d^{\rm l}/d^{\rm s} = 1.5\}$; and (d) Lower top-wall velocity case $\{\ell/\bar{d}_0=18,\,\tilde{v}_{\rm w}=0.01 \, c^{\rm l}_0=0.50, \, d^{\rm l}/d^{\rm s} = 1.5\}$. For each case, the first column shows  spatiotemporal contours of the evolution of $c^{\rm l}$ measured in DEM simulations. The second column shows comparisons of the DEM simulations (dashed gray lines) and continuum model predictions (solid black lines) of the $c^{\rm l}$ field at four different time instants during the segregation process: $\tilde{t}  = t/(\ell/v_{\rm w}) = 2 \times 10^1,\, 1 \times 10^2,\, 3 \times 10^2$, and $6 \times 10^2 $ in (a) and (b) and $\tilde{t}  = 1 \times 10^1,\, 5 \times 10^1,\, 1.5 \times 10^2$, and $3 \times 10^2 $ in (c) and (d) in the sequence of top left, top right, bottom left, bottom right. The third column shows comparisons of the steady-state, normalized velocity profiles from DEM simulations and continuum model predictions, corresponding to the last time instant listed above for the concentration field for each case.}
\label{fig:SegP_PSwG_transient_comparions_spheres}
\end{figure}

We compare continuum model predictions against corresponding DEM simulation results for bidisperse spheres using the parameter set given in \eqref{mater_param_seg_spheres}. To broadly exercise the model, we consider four different cases of planar shear flow with gravity: (1) the base case $\{\ell/\bar{d}_0=18,\,\tilde{v}_{\rm w}=0.02, \, c^{\rm l}_0=0.50, \, d^{\rm l}/d^{\rm s} = 1.5\}$, (2) a more large grains case $\{\ell/\bar{d}_0=18,\,\tilde{v}_{\rm w}=0.02, \, c^{\rm l}_0=0.75, \, d^{\rm l}/d^{\rm s} = 1.5\}$, (3) a larger loading length scale case $\{\ell/\bar{d}_0=36,\,\tilde{v}_{\rm w}=0.01 \, c^{\rm l}_0=0.50, \, d^{\rm l}/d^{\rm s} = 1.5\}$, and (4) a lower top-wall velocity case $\{\ell/\bar{d}_0=18,\,\tilde{v}_{\rm w}=0.01 \, c^{\rm l}_0=0.50, \, d^{\rm l}/d^{\rm s} = 1.5\}$, and results for these cases are shown in Figs.~\ref{fig:SegP_PSwG_transient_comparions_spheres}(a), (b), (c), and (d), respectively. The first column shows the spatiotemporal evolution of the concentration field $c^{\rm l}$ measured from the DEM simulations. In the second column, we compare snapshots of the concentration field $c^{\rm l}$ predicted by the continuum model (solid black lines) with the corresponding coarse-grained DEM fields (dashed gray lines) at four different time instants---$\tilde{t} = t/(\ell/v_{\rm w}) = 2 \times 10^1,\,1\times 10^2,\, 3 \times 10^2$, and $6 \times 10^2$ in Figs.~\ref{fig:SegP_PSwG_transient_comparions_spheres} (a) and (b) and $\tilde{t} = 1 \times 10^1,\,5\times 10^1,\, 1.5 \times 10^2$, and $3 \times 10^2$ in Figs.~\ref{fig:SegP_PSwG_transient_comparions_spheres} (c) and (d)---as indicated by the horizontal lines on the contour plots in the first column for each case. Finally, the steady-state normalized velocity fields (corresponding to the last time instant listed above for each case) obtained from DEM simulations and the corresponding continuum model predictions are shown in the last column of Fig.~\ref{fig:SegP_PSwG_transient_comparions_spheres}. The continuum model is able to capture the decaying velocity fields in all cases as well as predict the transient evolution of the segregation process.  

At this stage, it is instructive to contrast features of planar shear flow with gravity with inclined plane flow to emphasize why this flow configuration represents a non-trivial validation test for a coupled model for size-segregation and flow. At a high level, planar shear flow with gravity is boundary-driven, while inclined plane flow is gravity-driven. This difference in the manner in which flow is driven manifests in stark differences in the consequent flow fields. Namely, the Bagnold-like profile of inclined plane flow involves only modest strain-rate gradients, leading to segregation phenomenology that is dominated by pressure gradients. Moreover, the rheological behavior of a dense granular medium in inclined plane flow is dominated by local grain-inertia effects that are captured by the local inertial, or $\mu(I)$, rheology. This is in contrast to planar shear flow with gravity, in which the flow fields decay rapidly with the distance from the moving, top wall. While local grain-inertia effects contribute in the flowing region under the top wall, nonlocal, cooperative effects also contribute and become increasingly dominant with the distance from the top wall, as the flow field transitions into the quasi-static, creeping region. Therefore, utilizing a nonlocal rheology, such as the generalized NGF model for bidisperse flows, is crucial to accurately capture the velocity fields in planar shear flow with gravity. It is also notable that the magnitudes of the strain-rate gradients are comparatively greater in planar shear flow with gravity than in inclined plane flow, and the comparative contributions of the two mechanisms of segregation, i.e., pressure gradients and shear-strain-rate gradients, are unclear. We will return to this point in Section~\ref{sec:discussion}. Due to the distinctly different flow fields and the relative directions of the two segregation fluxes, i.e., competitive versus cooperative, inclined plane flow and planar shear flow with gravity represent markedly different flow configurations, so the observation that the coupled continuum model is capable of capturing segregation dynamics and flow in both configurations with a single set of parameters is encouraging.

\begin{figure}[t!]
\begin{center}
\includegraphics[width=0.79\columnwidth]{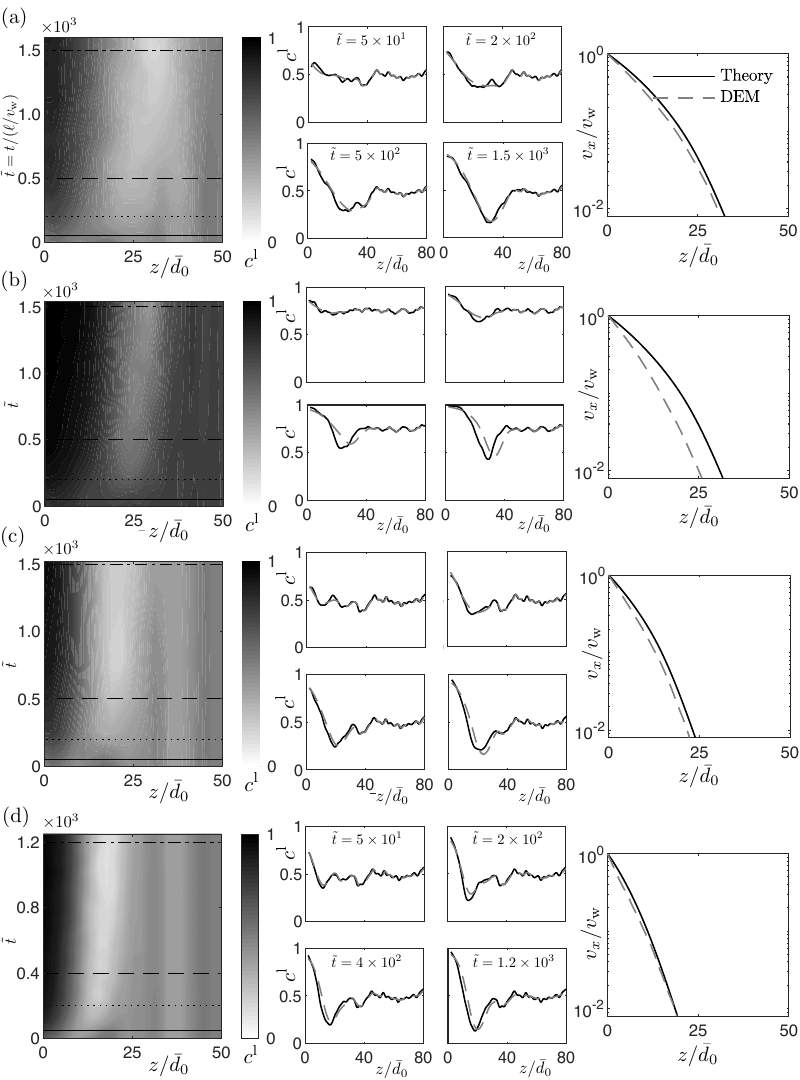}
\end{center}
\caption{Comparisons of continuum model predictions with corresponding DEM simulation results for the transient evolution of the segregation dynamics for four different cases of planar shear flow with gravity of bidisperse disks: (a) Base case $\{\ell/\bar{d}_0=60,\,\tilde{v}_{\rm w}=0.015, \, c^{\rm l}_0=0.50, \, d^{\rm l}/d^{\rm s} = 1.5\}$; (b) More large grains case $\{\ell/\bar{d}_0=60,\,\tilde{v}_{\rm w}=0.015, \, c^{\rm l}_0=0.75, \, d^{\rm l}/d^{\rm s} = 1.5\}$; (c) Smaller loading length scale case $\{\ell/\bar{d}_0=40,\,\tilde{v}_{\rm w}=0.015 \, c^{\rm l}_0=0.50, \, d^{\rm l}/d^{\rm s} = 1.5\}$; and (d) Lower top-wall velocity case $\{\ell/\bar{d}_0=60,\,\tilde{v}_{\rm w}=0.001 \, c^{\rm l}_0=0.50, \, d^{\rm l}/d^{\rm s} = 1.5\}$. Results are organized as described in the caption of Fig.~\ref{fig:SegP_PSwG_transient_comparions_spheres}. The time instants shown in the second column correspond to $\tilde{t}  = t/(\ell/v_{\rm w}) = 5 \times 10^1,\, 2 \times 10^2,\, 5 \times 10^2$, and $1.5 \times 10^3 $ in (a), (b) and (c) and $\tilde{t}  = 5 \times 10^1,\, 2 \times 10^2,\, 4 \times 10^2$, and $1.2 \times 10^3$ in (d).}
\label{fig:SegP_PSwG_transient_comparions_disks}
\end{figure}

Finally, we also test the performance of the model in planar shear flow with gravity of bidisperse disks. The material parameter set used to obtain continuum model predictions is given by \eqref{mater_param_seg_disks}. As discussed in Section~\ref{sec:IPF_transient_comparisons} for inclined plane flow, since normal stress differences are minimal in dense flows of disks, we do not use information from the coarse-grained DEM stress fields in our continuum simulations, instead using the analytical expressions for the stress ratio and pressure fields obtained from a static force balance. Again, we consider four different cases of planar shear flow with gravity---(1) the base case $\{\ell/\bar{d}_0=60,\,\tilde{v}_{\rm w}=0.015, \, c^{\rm l}_0=0.50, \, d^{\rm l}/d^{\rm s} = 1.5\}$, (2) a more large grains case $\{\ell/\bar{d}_0=60,\,\tilde{v}_{\rm w}=0.015, \, c^{\rm l}_0=0.75, \, d^{\rm l}/d^{\rm s} = 1.5\}$, (3) a smaller loading length scale case $\{\ell/\bar{d}_0=40,\,\tilde{v}_{\rm w}=0.015 \, c^{\rm l}_0=0.50, \, d^{\rm l}/d^{\rm s} = 1.5\}$, and (4) a lower top-wall velocity case $\{\ell/\bar{d}_0=60,\,\tilde{v}_{\rm w}=0.001 \, c^{\rm l}_0=0.50, \, d^{\rm l}/d^{\rm s} = 1.5\}$, and results for these cases are shown in Figs.~\ref{fig:SegP_PSwG_transient_comparions_disks}(a), (b), (c), and (d), respectively. In DEM simulations of planar shear flow with gravity of disks, the dimensions of the simulation domain are $H=120\bar{d}_0$ and  $L=60\bar{d}_0$ in all cases, both of which we have confirmed to be large enough to not affect the flow fields or the segregation dynamics. The first column shows contour plots of the spatiotemporal evolution of the $c^{\rm l}$ field measured from DEM simulations. The second column shows comparisons of the $c^{\rm l}$ field predicted by the continuum model with corresponding DEM results at four different time instants---$\tilde{t} = t/(\ell/v_{\rm w}) = 5 \times 10^1,\,2\times 10^2,\, 5 \times 10^2$, and $1.5 \times 10^3$ in Figs.~\ref{fig:SegP_PSwG_transient_comparions_disks}(a), (b), and (c) and $\tilde{t} = 5 \times 10^1,\,2\times 10^2,\, 4 \times 10^2$, and $1.2 \times 10^3$ in Fig.~\ref{fig:SegP_PSwG_transient_comparions_disks}(d)---as indicated by the horizontal lines on the contour plots in the first column. The third column shows the steady-state velocity field comparisons between the continuum model predictions and the DEM results. Again, the model does a reasonable job in predicting the segregation dynamics and the decaying flow fields across different cases of planar shear flow with gravity. Of note, in the case of the lower top-wall velocity, the entire flow is in the quasi-static, creeping regime, in which nonlocal effects are dominant. Therefore, the local inertial rheology could not capture any aspect of this flow field, and using the NGF model is crucial in this case to predict the flow field and hence the evolution of the $c^{\rm l}$ field. Overall, the coupled continuum model is capable of predicting the dynamics of the large-grain concentration field and the steady-state velocity field simultaneously across different loading conditions, initial conditions, and flow geometries for dense, bidisperse flows of both spheres and disks.

\section{Discussion} \label{sec:discussion}
\begin{figure}[!t]
\begin{center}
\includegraphics[width=0.81\columnwidth]{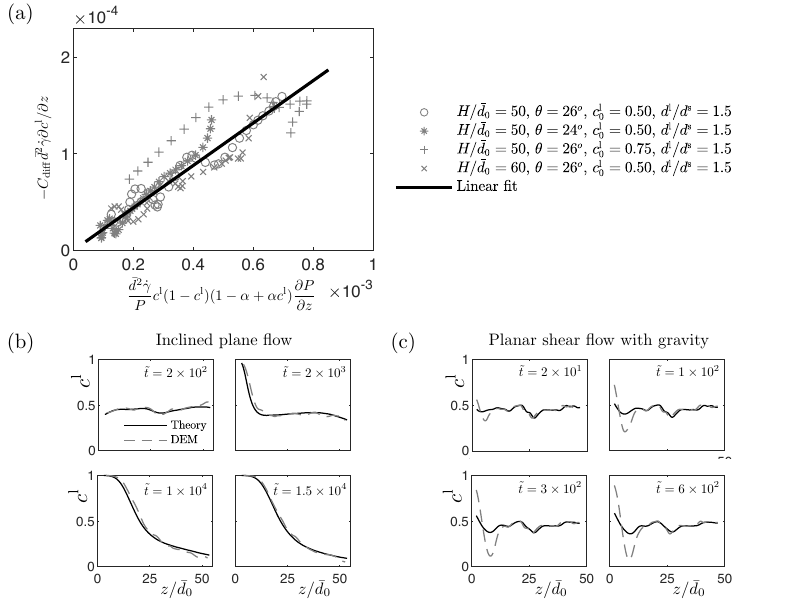}
\end{center}
\caption{(a) Collapse of $-C_{\rm diff} \bar{d}^2\dot{\gamma} ({\partial c^{\rm l}}/{\partial z})$ versus $({\bar{d}^2 \dot{\gamma}}/{P}) c^{\rm l}(1-c^{\rm l})(1 - \alpha + \alpha c^{\rm l}) ({\partial P}/{\partial z})$ for several cases of inclined plane flow of bidisperse spheres. Symbols represent coarse-grained, steady-state DEM field data, and the solid line is the best linear fit using $C^{\rm P}_{\rm seg} = 0.22$. Comparisons of continuum model predictions with corresponding DEM simulation results for the $c^{\rm l}$ field at four different time instants during the segregation process for (b) the base case of inclined plane flow of spheres and (c) the base case of planar shear flow with gravity of spheres but taking $C^{\rm S}_{\rm seg}=0$ and $C^{\rm P}_{\rm seg} = 0.22$ in the continuum model. Dashed gray lines represent the DEM simulations and solid black lines represent the continuum model predictions.}
  \label{fig:fig8}
\end{figure}

A natural question is whether it is essential to account for both shear-strain-rate-gradient-driven segregation and pressure-gradient-driven segregation in \eqref{eq:flux_decomp} to capture the dynamics of the large-grain concentration field in both inclined plane flow and planar shear flow with gravity using a single set of material parameters. To address this question, we first suppress the flux associated with shear-strain-rate gradients by taking $C^{\rm S}_{\rm seg} = 0$ and test whether the pressure-gradient-driven flux is sufficient to capture the segregation dynamics on its own. We start by reestimating the parameter $C^{\rm P}_{\rm seg}$ for the situation when $C^{\rm S}_{\rm seg} = 0$. We return to the steady-state DEM field data for the four cases of inclined plane flow of spheres described in Section~\ref{sec:PGDS}, and as suggested by \eqref{eq:flux_balance}, we plot $-C_{\rm diff}\bar{d}^2\dot\gamma(\partial c^{\rm l}/\partial z)$ versus $({\bar{d}^2 \dot{\gamma}}/{P}) c^{\rm l}(1-c^{\rm l})(1 - \alpha + \alpha c^{\rm l}) ({\partial P}/{\partial z})$ for $\alpha=0.4$ in Fig.~\ref{fig:fig8}(a), where each data point represents a unique $z$-position. It is evident that the collapse of the data is not as strong in Fig.~\ref{fig:fig8}(a) as it is in Fig.~\ref{fig:flux_balance}(a), where the shear-strain-rate-gradient-driven flux is included. Nevertheless, a linear trend is observed, and we reestimate $C^{\rm P}_{\rm seg} = 0. 22$ from the slope of the best linear fit shown in Fig.~\ref{fig:fig8}(a). This value is lower than the previously determined value of 0.51 due to the omission of the shear-strain-rate-gradient-driven flux, which acts in the opposite direction of the pressure-gradient-driven flux. Next, we compare the transient evolution of the $c^{\rm l}$ field for the base case of inclined plane flow of spheres due to the pressure-gradient-driven flux solely, i.e., using $C^{\rm S}_{\rm seg} = 0$ and $C^{\rm P}_{\rm seg} = 0. 22$. Otherwise, the solution procedure is the same as described in Section~\ref{sec:IPF_transient_comparisons}. The results are shown in Fig.~\ref{fig:fig8}(b) for the $c^{\rm l}$ field at four different time instants during the segregation process, where dashed gray lines represent the DEM simulations and solid black lines represent the continuum model predictions. The comparisons look reasonable in spite of not accounting for the shear-strain-rate-gradient-driven flux and are comparable to the results shown in Fig.~\ref{fig:SegP_IPF_transient_comparisons_spheres}(a) for the same case when both flux contributions are included. This confirms that the pressure-gradient-driven flux is dominant in inclined plane flow but might lead one to believe that the shear-strain-rate-gradient-driven flux is always negligible when pressure gradients are present. Therefore, we return to the base case of planar shear flow with gravity of spheres and calculate predictions of the continuum model for the transient evolution of the $c^{\rm l}$ field using $C^{\rm S}_{\rm seg} = 0$ and $C^{\rm P}_{\rm seg} = 0. 22$. The results are shown in Fig.~\ref{fig:fig8}(c) for four time instants during the segregation process, and for these parameters, the continuum model grossly under-predicts the extent of segregation when compared to the results of Fig.~\ref{fig:SegP_PSwG_transient_comparions_spheres}(a) when both flux contributions are included. Therefore, in planar shear flow with gravity, pressure-gradient-driven segregation is a secondary effect with the segregation process being primarily driven by shear-strain-rate gradients, and it is not possible to capture the segregation dynamics in both flow configurations when neglecting the shear-strain-rate-gradient-driven flux. 

\begin{figure}[!t]
\begin{center}
\includegraphics[width=0.62\columnwidth]{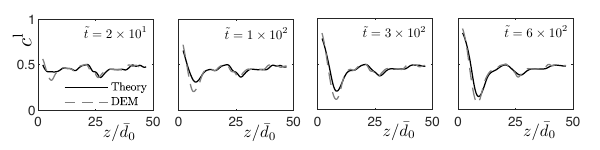}
\end{center}
\caption{Comparisons of continuum model predictions with corresponding DEM simulation results for the base case of planar shear flow with gravity  of spheres but taking $C^{\rm S}_{\rm seg}=0.08$ and $C^{\rm P}_{\rm seg} = 0$ in the continuum model. Dashed gray lines represent the DEM simulations and solid black lines represent the continuum model predictions.}
\label{fig:fig11}
\end{figure}

On the other hand, it is also worth considering the scenario in which the pressure-gradient-driven flux is neglected, i.e., $C^{\rm P}_{\rm seg} = 0$. In this situation, we do not reestimate $C^{\rm S}_{\rm seg}$ since it has been independently determined in isolation in our prior work \citep{liu_singh_arxiv}. As discussed in Section~\ref{sec:PGDS}, in inclined plane flow, the strain rate is greatest at the bottom of the layer, so the shear-strain-rate-gradient-driven flux drives the large grains towards the bottom of the layer. Therefore, neglecting the pressure-gradient-driven flux would lead to the model predicting that segregation evolves in the opposite direction from what is observed in DEM simulations. We also consider planar shear flow with gravity in the absence of the pressure-gradient-driven flux and revisit the base case for spheres. Results for four time instants during the segregation process are shown in Fig.~\ref{fig:fig11}, using $C^{\rm S}_{\rm seg}=0.08$ and $C^{\rm P}_{\rm seg} = 0$ in the continuum model. The comparisons appear reasonable, but the predictions of the model still lag behind the corresponding DEM results. Comparing the continuum model predictions in Fig.~\ref{fig:fig11} with those in Fig.~\ref{fig:SegP_PSwG_transient_comparions_spheres}(a), it is evident that accounting for both driving mechanisms results in the most accurate predictions. In conclusion, although the shear-strain-rate-gradient-driven flux is the primary driver in planar shear flow with gravity, both mechanisms should be included to accurately capture the segregation dynamics, and more broadly, it is crucial to account for both mechanisms to capture segregation dynamics across different flow geometries with a single set of parameters.

\section{Concluding remarks} \label{sec:conclusion}
In this paper, we studied coupled flow and size-segregation in dense, bidisperse granular systems of frictional spheres and disks in scenarios when both pressure gradients and shear-strain-rate gradients are present, and we developed a phenomenological continuum model that can simultaneously capture the segregation dynamics and velocity fields. The continuum model integrates a constitutive equation for the pressure-gradient-driven segregation flux, based on the works of \citet{gajjar2014asymmetric} and \citet{trewhela2021experimental}, with the coupled model for size-segregation and flow in the absence of pressure gradients developed in our prior work \citep{liu_singh_arxiv}. The complete, coupled model for bidisperse granular systems accounts for pressure-gradient-driven and shear-strain-rate-gradient-driven segregation fluxes, granular diffusion, and nonlocal rheological behavior spanning the quasi-static and dense inertial flow regimes. The segregation model involves four dimensionless material parameters $\{C_{\rm diff}, C^{\rm S}_{\rm seg}, C^{\rm P}_{\rm seg},\alpha\}$, and the parameters associated with the pressure-gradient-driven flux $\{C^{\rm P}_{\rm seg},\alpha\}$ were determined in this work for both frictional spheres and disks with a size ratio of $d^{\rm l}/d^{\rm s}=1.5$, based on steady-state DEM data in inclined plane flow. The coupled model has been tested in two flow configurations involving pressure gradients---namely, (1) inclined plane flow and (2) planar shear with gravity flow---and the coupled continuum model does an excellent job capturing the transient evolution of the segregation fields as well as the steady-state velocity fields across several variants of both flow configurations, using one set of parameters for spheres and another for disks. 

There remain a number of important directions for future work---three of which are highlighted here. First, in this paper, we focused on a single grain-size ratio, $d^{\rm l}/d^{\rm s}=1.5$; however, many works in the literature \citep[e.g.,][]{schlick2015,tunuguntla2017,trewhela2021experimental} have shown that the rate of pressure-gradient-driven size-segregation depends on the grain-size ratio. We have corroborated this with several of our own tests, in which steady-state DEM field data from inclined plane flow of bidisperse granular systems with grain-size ratios not equal to 1.5 do not collapse with the data of Fig.~\ref{fig:flux_balance}. These observations indicate that the constitutive equation for the pressure-gradient-driven segregation flux \eqref{eq:segP_eqn} needs to be generalized to account for dependence on $d^{\rm l}/d^{\rm s}$, and we expect that the explicit dependence determined in \citet{trewhela2021experimental} may be leveraged to this end. We note that this dependence of the pressure-gradient-driven segregation flux on the grain-size ratio is in contrast to the shear-strain-rate-gradient-driven segregation flux, where we did not observe grain-size-ratio dependence over a similar range of modest grain-size ratios \citep{liu_singh_arxiv}. Second, the flow geometries considered in this paper result in planar, shearing flows, in which pressure gradients lie within the plane of shearing. In many flows, e.g., annular shear flow with gravity, pressure gradients act perpendicular to the plane of shearing, and it remains to test whether the flux constitutive equations utilized in this paper continue to be valid in such scenarios or require modification. Finally, a robust numerical toolkit needs to be developed to solve the coupled system of governing equations in more complex flow geometries, where the stress field cannot be straightforwardly deduced. 

\section*{Acknowledgements}
This work was supported by funds from NSF-CBET-1552556.

\appendix
\section{Rheological constitutive equations} \label{app:flow_model}
\subsection{Local inertial rheology}\label{app:local}
The local inertial, or $\mu(I)$, rheology has been widely used to describe dense granular flows \citep{midi2004, jop2005, dacruz2005} and follows from straightforward dimensional arguments. Here, we summarize the generalization of the local inertial rheology for a dense, bidisperse system of dry, stiff grains with average grain diameter $\bar{d}$ (defined in Section~\ref{subsec_mixt_seg}) and grain-material mass density $\rho_{\rm s}$, as introduced in \citet{rognon07} and \citet{tripathi11}. The local inertial rheology relates the stress ratio $\mu$, the equivalent shear strain rate $\dot\gamma$, and the pressure $P$ (each of which are tensor invariant quantities defined in either Section~\ref{subsec_kine} or \ref{subsec_eom}) through the dimensionless relationship $\mu = \mu_{\rm loc}(I)$, where $I = \dot{\gamma}\sqrt{\bar{d}^2 \rho_{\rm s} / P}$ is the inertial number.

In this work, we utilize the following nonlinear functional form of \citet{jop2005} for dense systems of bidisperse spheres:
\begin{equation}\label{eq:mu_I_spheres}
\mu_{\rm loc}(I) = \mu_{\rm s} + \frac{\mu_2 - \mu_{\rm s}}{I_0/I +1}, 
\end{equation}
where $\mu_{\rm s}$ is the stress ratio corresponding to the static yield condition, $\mu_2$ corresponds to the stress ratio in the limit $I \to \infty$, and $I_0$ is a dimensionless parameter which characterizes the nonlinear, rate-dependent response. The rheological material parameters for dense, frictional spheres have been determined from DEM simulations to be $\{\mu_{\rm s}=0.37, \mu_2=0.95, I_0=0.58\}$ for the monodisperse case \citep{liu_singh_arxiv,zhang2017}, and as discussed in \citet{liu_singh_arxiv}, the parameters may continue to be used to describe the rheology of dense, frictional spheres in the bidisperse case.

For dense systems of bidisperse disks, we utilize the following linear functional form of \citet{dacruz2005}:
\begin{equation}\label{eq:mu_I_disks}
\mu_{\rm loc}(I) = \mu_{\rm s} + bI,
\end{equation}
where $\mu_{\rm s}$ plays the same role as in \eqref{eq:mu_I_spheres} and $b$ is a dimensionless parameter, which characterizes the slope of the functional relationship. The rheological parameters for dense, frictional disks, which are applicable in both the monodisperse and bidisperse cases, have been determined from DEM simulations to be $\{\mu_{\rm s}=0.272, b=1.168\}$ \citep{liu2018,liu_singh_arxiv}.

\subsection{Nonlocal granular fluidity model}
While the local inertial rheology can capture homogeneous flows well, it fails to capture many aspects of flow fields with spatial inhomogeneity. In order to capture inhomogeneous flows and account for size-dependent effects, several nonlocal models have been developed \citep{kamrin2019non}. In this work, we utilize the nonlocal granular fluidity (NGF) model which has been developed and broadly tested for dense, monodisperse granular systems \citep[e.g.,][]{kamrin2012, henann2013} and recently extended to dense, bidisperse granular systems \citep{liu_singh_arxiv}. The model is summarized as follows. A positive, scalar field quantity---called the granular fluidity $g$---is introduced, which relates the stress state to the strain rate by means of two constitutive equations. First, the flow rule relates the Cauchy stress tensor $\sigma_{ij}$, the strain-rate tensor $D_{ij}$, and the granular fluidity $g$ through $\sigma_{ij} = -P\delta_{ij} + 2({P}/{g})D_{ij}$, where we have made the common idealization that the Cauchy stress deviator and the strain-rate tensor are codirectional \citep{rycroft2009}. Taking the magnitude of the deviatoric part of the flow rule and rearranging gives the scalar form of the flow rule, 
\begin{equation}\label{eq:flow_rule}
\dot{\gamma} = g \mu,
\end{equation}
which is the form that we utilize in Section~\ref{sec:validation}. Second, the granular fluidity field $g$ is governed by the nonlocal rheology:
\begin{eqnarray}\label{eq:nl_rheology}
g = g_{\rm loc} (\mu, P) + \xi^2(\mu) \frac{\partial^2 g}{\partial x_i \partial x_i},
\end{eqnarray}
where $g_{\rm loc}(\mu,P)$ is the local fluidity function and $\xi(\mu)$ is the stress-dependent cooperativity length. Using the functional form of the local inertial rheology given in \eqref{eq:mu_I_spheres} and the definition of the inertial number for bidisperse systems leads to the following local fluidity function for the case of dense, bidisperse spheres:
\begin{equation}\label{eq:localg_bi_seg}
g_{\rm loc}(\mu,P) =  \left\{\begin{array}{cl}I_0 \sqrt{\dfrac{P}{\bar{d}^2\rho_{\rm s}}} \, \dfrac{(\mu-\mu_{\rm s})}{\mu (\mu_2 - \mu)}  &{\rm if}\ \mu>\mu_{\rm s},\\[4pt]
0 &{\rm if}\ \mu\le\mu_{\rm s}.\end{array}\right.
\end{equation}
Similarly, using \eqref{eq:mu_I_disks}, the local fluidity function for the case of dense, bidisperse disks is
\begin{equation}\label{eq:localg_bi_seg_disks}
g_{\rm loc}(\mu,P) =  \left\{\begin{array}{cl} \sqrt{\dfrac{P}{\bar{d}^2\rho_{\rm s}}} \, \dfrac{(\mu-\mu_{\rm s})}{b\mu}  &{\rm if}\ \mu>\mu_{\rm s},\\[4pt]
0 &{\rm if}\ \mu\le\mu_{\rm s}.\end{array}\right.
\end{equation}
Finally, the functional forms for the cooperativity length corresponding to \eqref{eq:localg_bi_seg} and \eqref{eq:localg_bi_seg_disks} are
\begin{equation}\label{eq: cooperativity_bi_seg}
\xi(\mu) = A\bar{d}\sqrt{\frac{(\mu_2 - \mu)}{(\mu_2 - \mu_{\rm s})|\mu - \mu_{\rm {\rm s}}|}} \qquad\text{and}\qquad \xi (\mu)=  \frac{A\bar{d}}{\sqrt{|\mu-\mu_{\rm s}|}},
\end{equation}
where $A$ is a dimensionless parameter called the nonlocal amplitude. For both monodisperse and bidisperse systems, the nonlocal amplitude has been determined from DEM simulations to be $A=0.43$ for dense, frictional spheres \citep{zhang2017,liu_singh_arxiv} and $A=0.90$ for dense, frictional disks \citep{liu2018,liu_singh_arxiv}. In the NGF model, nonlocal effects are accounted through the additional scalar state variable $g$, which is governed by the differential relation \eqref{eq:nl_rheology} that involves one new dimensionless material parameter $A$. Using this approach, the NGF model can capture the significant deviation from a local, one-to-one constitutive relationship $\mu = \mu_{\rm loc}(I)$ that is observed in experiments and DEM simulations of inhomogeneous, dense granular flow.

\section{Discrete element method simulations and averaging methods}
\subsection{Simulated granular systems}\label{app:DEM}
We consider three-dimensional granular systems consisting of dense, bidisperse mixtures of spheres and two-dimensional granular systems consisting of dense, bidisperse mixtures of disks. For both spheres and disks, the mean large-grain diameter is $d^{\rm l} = 3$\,mm, and the mean small-grain diameter is $d^{\rm s} = 2$\,mm. The grain-size ratio $d^{\rm l}/d^{\rm s} = 1.5$ is held fixed throughout this study. To prevent crystallization, we introduce a polydispersity of $\pm 10 \%$ to the respective mean diameters of both  species. The grain-material volume-density is $\rho_{\rm s} = 2450\, {\rm kg/m^3}$ for spheres, and for disks, the grain-material area-density is $\rho_{\rm s} = 3.26\, {\rm kg/m^2}$. The model for the grain-grain interaction force is given through a contact law that accounts for elasticity, damping, and sliding friction, which is widely used in the literature \citep[e.g.,][]{dacruz2005,koval2009,kamrin2014,zhang2017}. The details of this contact law are not repeated here, but we highlight the parameters that fully describe the interaction properties: (1) $k_{\rm n}$ the normal contact stiffness, (2) $k_{\rm t}$ the tangential contact stiffness, (3) the coefficient of restitution for binary collisions $e$, and (4) the inter-particle friction coefficient $\mu_{\rm surf}$. The normal contact stiffness is taken to be sufficiently large compared to the characteristic confining pressure, so that the grains are nearly rigid, i.e., $k_{\rm n}/P\bar{d}_0 > 10^4$ for spheres and $k_{\rm n}/P > 10^4$ for disks. We take $k_{\rm t}/ k_{\rm n}=1/2$ and $e=0.1$, but these parameters have negligible effects on flow and size-segregation in the rigid-grain limit. In this regime, the only interaction parameter that significantly affects flow and size-segregation phenomenology is $\mu_{\rm surf}$, which we maintain as $\mu_{\rm surf}=0.4$ throughout. Lastly, numerical integration of the equations of motion for each grain is performed using the open-source software LAMMPS \citep{lammps}, and the time step for numerical integration is chosen to be sufficiently small, compared to the binary collision time, to ensure the stability and accuracy of simulation results.   

\subsection{Averaging methods}\label{app:coarse_grain}
We utilize a bin-based approach for spatial averaging of a given snapshot of DEM data at time $t$, which is described here for the case of spheres in three dimensions. Consider a rectangular-cuboidal bin of width $\Delta$, centered at a position $z$ and spanning the simulation domain along the $x$- and $y$-directions. At time $t$, we assign each grain $i$ intersected by the bin a weight $V_i$, which is equal to the volume of grain $i$ inside the bin. As in \citet{tunuguntla2017}, we denote the sets of large and small grains intersected by the bin as ${\cal F}^{\rm l}$ and ${\cal F}^{\rm s}$, respectively, so that the set of all grains intersected by the bin is ${\cal F} = {\cal F}^{\rm l} \cup {\cal F}^{\rm s}$ with ${\cal F}^{\rm l} \cap {\cal F}^{\rm s} = \varnothing$. The instantaneous solid volume fraction field for species $\nu = {\rm l},{\rm s}$ is $\phi^{\nu}(z,t) = (\sum_{i \in {\cal F}^\nu} V_i)/V$, where $V$ is the total volume of the bin, and the concentration field for species $\nu$ is $c^{\nu}(z,t) = \phi^{\nu}(z,t)/\phi(z,t)$ with $\phi(z,t)=\phi^{\rm l}(z,t)+\phi^{\rm s}(z,t)$. The instantaneous velocity field of the mixture at position $z$ and time $t$ is $\boldsymbol{v}(z,t) = (\sum_{i \in {\cal F}} V_i\boldsymbol{v}_i(t))/(\sum_{i \in {\cal F}} V_i)$, where $\boldsymbol{v}_i(t)$ is the instantaneous velocity of each grain $i$. The instantaneous stress tensor associated with each grain $i$ is $\boldsymbol{\sigma}_i(t) = (\sum_{j\ne i} \boldsymbol{r}_{ij}\otimes \boldsymbol{f}_{ij})/(\pi d_i^3/6)$, where $\boldsymbol{r}_{ij}$ is the position vector from the center of grain $i$ to the center of grain $j$, $\boldsymbol{f}_{ij}$ is the contact force applied on grain $i$ by grain $j$, and $d_i$ is the diameter of grain $i$. The instantaneous stress field of the mixture at position $z$ and time $t$ is then $\boldsymbol{\sigma}(z,t) = (\sum_{i \in {\cal F}} V_i\boldsymbol{\sigma}_i(t))/V$. We note that in the limit $\Delta \to 0 $, the bins reduce to slices with zero width, spanning the simulation domain along the $x$- and $y$-directions. Slice-based approaches of this type have been successfully used in the literature to obtain coarse-grained velocity and stress fields \citep[e.g.,][]{dacruz2005,koval2009,zhang2017}. However, concentration fields obtained using slice-based coarse-graining exhibit spatial fluctuations due to particle layering near walls \citep{weinhart2013}. Here, we are careful to choose a bin width that is sufficiently large to avoid layering effects but not large enough to over-smooth the DEM data. Throughout this study, as in \citet{liu_singh_arxiv}, we take a bin width of $\Delta = 4 \bar{d}_0$ and a spatial resolution of about $0.1 \bar{d}_0$. One exception is for the velocity fields in planar shear flow with gravity, where we use a slice-based approach, i.e., $\Delta \to 0$. This is done to control for any wall slip and precisely estimate the velocity of the first layer of flowing grains beneath the top wall in the DEM simulations, which then corresponds to the velocity $v_{\rm w}$ quoted for each case in Section~\ref{sec:PSwG_transient_comparisons} and matched in the corresponding continuum simulations. The spatial averaging method for the case of disks in two dimensions follows an analogous, bin-based approach, using area-based weights rather than volume-based weights, and has been described in detail in our prior work \citep{liu_singh_arxiv}, so it is not repeated here. When plotting spatiotemporal contours of the concentration fields, profiles of the concentration fields, and profiles of the velocity fields (e.g., Fig.~\ref{fig:SegP_IPF_transient_comparisons_spheres}), we truncate the coarse-grained DEM data from bins centered within about $3\bar{d}_0$ of both the top and bottom boundaries. For the steady-state collapses of Fig.~\ref{fig:flux_balance}, we use a more conservative criteria to eliminate any potential boundary effects and truncate DEM data from bins centered within about $6\bar{d}_0$ of the boundaries. When calculating the quantities appearing in the collapses of Fig.~\ref{fig:flux_balance}, the instantaneous fields are spatially smoothed and differentiated using a kernel function. We have tested both a cutoff Gaussian kernel function and Lucy's quartic kernel function \citep{weinhart2013,tunuguntla2016} and a range of kernel function widths to ensure that the reported results in this paper are insensitive to these choices. To then obtain the relevant steady-state field quantities in \eqref{eq:flux_balance}, we consider a time window within the steady-state regime and generate $N=1000$ snapshots of the instantaneous, smoothed fields and arithmetically average these fields over all snapshots to obtain fields that only depend on the spatial coordinate.

\end{document}